\RequirePackage{fix-cm}
\documentclass[twocolumn,epjc3]{svjour3}  
\smartqed  
\RequirePackage{graphicx}
\RequirePackage{physics}
\RequirePackage{latexsym}
\RequirePackage[numbers,sort&compress]{natbib}
\RequirePackage{amsfonts}
\RequirePackage[colorlinks,citecolor=blue,urlcolor=blue,linkcolor=blue]{hyperref}
\RequirePackage{amsmath}
\journalname{Eur. Phys. J. C}
\begin{document}

\title{\boldmath Probing Light Scalars and Vector-like Quarks at the High-Luminosity LHC}


\author{U. S. Qureshi \thanksref{e1,addr1}
        A. Flórez \thanksref{e2,addr2}
        \and
        A. Gurrola \thanksref{e3,addr1}
        \and 
        C. Rodriguez\thanksref{e4,addr2} \and 
         }

\thankstext{e1}{umar.sohail.qureshi@vanderbilt.edu}
\thankstext{e2}{ca.florez@uniandes.edu.co}
\thankstext{e3}{alfredo.gurrola@vanderbilt.edu}
\thankstext{e4}{c.rodriguez45@uniandes.edu.co}


\institute{Department of Physics and Astronomy, Vanderbilt University, Nashville, TN, USA \label{addr1}
           \and
           Physics Department, Universidad de los Andes, Bogotá, Colombia \label{addr2}
}

\date{\today}

\maketitle

\begin{abstract}
A model based on a $U(1)_{T^3_R}$ extension of the Standard Model can address the mass hierarchy between generations of fermions, explain thermal dark matter abundance, and the muon $g - 2$, $R_{(D)}$, and $R_{(D^*)}$ anomalies. The model contains a light scalar boson $\phi'$ and a heavy vector-like quark $\chi_\mathrm{u}$ that can be probed at CERN's Large Hadron Collider (LHC). We perform a phenomenology study on the production of $\phi'$ and ${\chi}_u$ particles from proton-proton $(\mathrm{pp})$ collisions at the LHC at $\sqrt{s}=13.6$ \textrm{TeV}, primarily through $g{-g}$ and $t{-\chi_\mathrm{u}}$ fusion. We work under an effective field theory approach, in which the $\chi_\mathrm{u}$ and $\phi'$ masses are free parameters. We perform a phenomenological analysis considering $\chi_\mathrm{u}$ final states to \textrm{b}-quarks, muons, and  neutrinos, and  $\phi'$ decays to $\mu^+\mu^-$. A machine learning algorithm is used to maximize the signal sensitivity, considering an integrated luminosity of $3000$ $\textrm{fb}^{-1}$. The proposed methodology can be a key mode for discovery over a large mass range, including low masses, traditionally considered difficult due to experimental constraints.
\end{abstract}











\section{Introduction}
\label{introduction}

The Standard Model (SM) of particle physics, despite its successful account of numerous experimental findings involving strong, electromagnetic, and weak interactions, confirmed by the CERN's Large Hadron Collider (LHC), is increasingly seen as a lower-energy vestige of a more comprehensive theory. This perspective arises from unresolved questions regarding the origins of dark matter, electroweak symmetry breaking scales, lepton flavor universality, the anomalous muon magnetic moment~\cite{PhysRevD.73.072003, g2cit,Davier2017,Davier2020,PhysRevLett.121.022003,PhysRevD.97.114025,PhysRevLett.124.132002,PhysRevD.100.076004}, discrepancies in the $R_{(D)}$ and $R_{(D^{*})}$ ratios from $\mathrm{b}-$meson decays~\cite{ BaBar:2012obs,BaBar:2013mob, Huschle:2015rga,LHCb:2015gmp,Aaij:2015yra,Sato:2016svk, Hirose:2016wfn, Aaij:2017uff, Hirose:2017dxl,LHCb:2017rln,Abdesselam:2019dgh,Belle:2019rba,LHCb:2023zxo}, as well as theoretical conundrums about whether gravity should be quantized, how gauge interactions can be unified, and how to explain divergences in the Higgs mass calculations. Furthermore, the SM offers no explanation for fermion family replication, nor for the lack of CP violation in the strong sector. These theoretical gaps, coupled with the experimental observation of phenomena such as neutrino masses, dark matter, and the baryon asymmetry in the Universe, which cannot be explained by the SM, reinforce the expectation for physics beyond the SM (BSM).

As a result, several theoretical models have been put forth to address the limitations of the SM over the past decade. Despite differing theoretical motivations and resulting implications, a common thread among these ideas is the introduction of new particles, that, depending on the model, might be probed via proton-proton $(\mathrm{pp})$ collisions at the  LHC. A myriad of ideas have been suggested to investigate BSM physics, driving a substantial amount of exploration at LHC. That research has significantly limited the scope of theories and established exclusionary boundaries extending to multi-\textrm{TeV} ranges for the masses of newly predicted particles within these theories \cite{ParticleDataGroup:2024cfk, CMS:2018iye, CMS:2016ucr, CMS:2016xbv, CMS:2016fxb, CMS:2017xcw, CMS:2015jsu}. Possible reasons for the absence of evidence could be attributed to new particle masses being at the threshold where they are too large to be produced at the LHC energies and likely with exceptionally low production rates. In the scenario where the masses of the new particles might be probed at the LHC but their production cross sections are small with respect to SM processes, a vast amount of data might be needed, together with advanced analysis techniques, to enhance the probability of detection. Alternatively, it is conceivable that new physics diverges from the conventional assumptions made in many BSM theories and the associated explorations. As a result, these new physics phenomena could remain hidden in processes that have not yet been thoroughly examined.

Minimal extensions to the SM, considering new $U(1)_{\chi}$ symmetry groups, are among the most studied BSM scenarios. For example, the  $U(1)_{T^3_R}$ symmetry, where families of right-handed fermions of the SM and possible extensions, such as right-handed neutrinos, are charged, was originally studied in the context of left-right symmetry models \cite{PatiSalam1974, MohapatraPati1975, SenjanovicMohapatra1975}. In these studies, $U(1)_{T^3_R}$ is identified as the subgroup of $SU(2)_R$ defined by its diagonal (electric-charge neutral) generator, $T^3_R$. In addition, it is often suggested that  $U(1)_{T^3_R}$ is a subspecies of a $U(1)_{B-L}$ symmetry since the breaking of the $U(1)_{B-L} \times U(1)_{T^3_R}$ leads to the $ U(1)_Y$ symmetry. This naturally motivates the presence of a massive and electrically neutral $\textrm{Z}'$ gauge boson~\cite{DiLuzio2018, Baker2019, Michaels:2020fzj, Dev:2021otb, Florez2023}. However, in the breaking of  $U(1)_{B-L} \times U(1)_{T^3_R} \rightarrow U(1)_Y$, it follows that the Higgs doublet $\mathrm{H}$, since it is a singlet of $U(1)_{B-L}$,  acquires its hypercharge by inheritance from a charge under $U(1)_{T^3_R}$. Consequently, the vacuum expectation value (VEV) of $\mathrm{H}$  couples both symmetry-breaking scales for $U(1)_Y$ and $U(1)_{T^3_R}$. Alternatively, these symmetry-breaking scales can be decoupled by adding an additional $U(1)_G$ group where fermions of the SM are singlets and $\mathrm{H}$ is not. Therefore, the hypercharge comes from $U(1)_G$ for the $\mathrm{H}$ and from $U(1)_{T^3_R}$ for fermions, \textit{i.e.} $Y=Q_{T^3_R}+\frac{1}{2}Q_{B-L} + Q_G$~\cite{Dutta:2022qvn}. Moreover, one can ask for scenarios where the hypercharge is not related to the $U(1)_{T^{3}_{R}}$ charge. 

Recently, theoretical and phenomenological efforts have emerged around scenarios where the low-energy gauge symmetry of the SM is extended by appending the Abelian gauge group $U(1)_{T^{3}_{R}}$, whose spontaneous symmetry-breaking is not linked to the electroweak one \cite{Dutta2019, Dutta2020, Dutta2020b,Dutta2022, PhysRevD.107.095019, Dutta2023}. In these scenarios, the gauge boson of $U(1)_{T^3_R}$ is associated with a massive dark photon $A'$ whose longitudinal mode arises from a Higgs--like mechanism involving a complex scalar field, $\phi$. This field is a singlet under the SM group, with its CP-odd component associated with the $A'$ mass and the CP-even giving rise to a dark Higgs, $\phi'$. To cancel gauge anomalies, a right-handed $\nu_R$ neutrino must be included for each generation of the SM that couples to $U(1)_{T^3_R}$. Furthermore, to correctly explain the origin of fermion masses in a UV-complete theory, a set of new vector-like quarks $(\chi_\mathrm{u}, \chi_d,\chi_\ell, \chi_\nu)$ must be included. These new particles are singlets under $U(1)_{T^3_R}$ and charged like SM right-handed fermions, as in the universal see-saw mechanism~\cite{Berezhiani, Chang1987, Davidson1987, Rajpoot1987, Babu1989, Babu1990}. 

\begin{figure}
    \centering
    \includegraphics[width=0.99\linewidth]{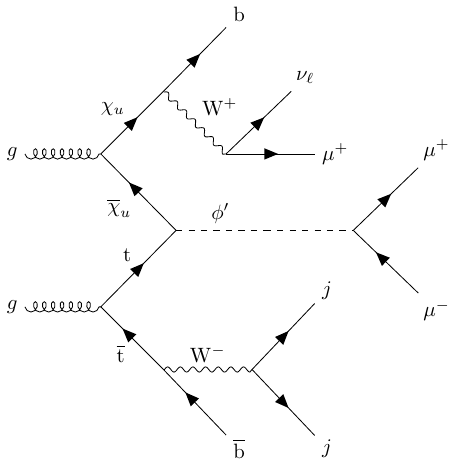}
    \caption{Representative Feynman diagram for the production of a $\phi'$ boson in association with a $\chi_\mathrm{u}$ vector-like quark through the fusion of a top quark and $\chi_\mathrm{u}$ vector-like quark. Once again, the $\phi'$ decays to a pair of muons, the top quark decays fully hadronically, and the $\chi_\mathrm{u}$ decays semi-leptonically to muons, neutrinos and $b$-jets.\label{fig:qqfusion}}
\end{figure}

\begin{figure}
    \centering
    \includegraphics[width=0.99\linewidth]{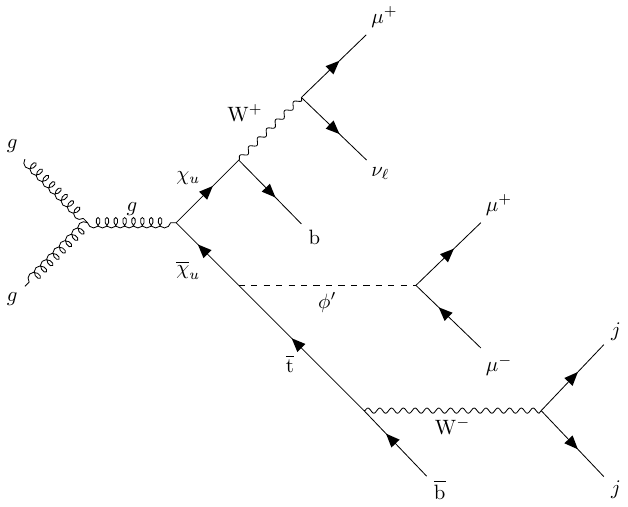}
    \caption{Representative Feynman diagram for the production of a $\phi'$ boson in association with a $\chi_\mathrm{u}$ vector-like quark through the fusion of a gluon pair from incoming protons. The $\phi'$ decays to a pair of muons, the top quark that decays fully hadronically, and the $\chi_\mathrm{u}$ decay semi-leptonically to muons, neutrinos and jets.\label{fig:ggfusion}}
\end{figure}

In this phenomenology study, we devise a LHC search strategy for the light \textrm{GeV}-scale scalar boson $\phi'$ produced in association with a heavy \textrm{TeV}-scale $\chi_\mathrm{u}$, the partner particle of the top quark, through a previously unexplored production and final state channel. Particularly, we explore the production of $\mathrm{pp}\to \mathrm{t}\chi_\mathrm{u} \phi'$, in contrast to $\mathrm{pp}\to \mathrm{T}\mathrm{T}\to \mathrm{t}\phi'\mathrm{t}\phi'$ with hadronic \cite{Bhardwaj_2022, Bhardwaj_2022_2, Bardhan_2023} di-photonic $\phi'$ \cite{Banerjee_2016, Alves_2024} decays. Due to the non-trivial $\chi - \mathrm{t} -\phi'$ coupling, processes where the final state includes $ \mathrm{t}\chi_\mathrm{u} \phi'$ are allowed in \textrm{pp} colliders through the ${\chi_\mathrm{u}}{- \mathrm{t}}$ fusion, see Figure~\ref{fig:qqfusion}. Since the $\chi_\mathrm{u}$ couples to SM quarks and gluons, it can be produced in large quantities. Furthermore, its energetic decay products can be detected alongside the $\phi'$ mediator particle that has significant transverse momentum. Therefore, if the $\phi'$ decays into SM particles that are observable in the detector's central region, this strategy can be very effective at reducing the SM background, and thus improve the long-term LHC discovery reach for heavy top partners and GeV-scale mediators, which are typically hard to detect using conventional methods at hadron colliders. Moreover, since 
it is possible to have $\chi_\mathrm{u} \to \mathrm{t}\,\phi'$ decays (and $\bar{\chi_\mathrm{u}} \to \bar{\mathrm{t}}\,\phi'$), the same $\mathrm{pp}\to \mathrm{t}\chi_\mathrm{u} \phi'$ state may arise from $\chi_\mathrm{u}\bar\chi_\mathrm{u}$ production diagrams with quantum chromodynamic (QCD) vertices, where one $\chi_\mathrm{u}$ decays to $\mathrm{t}\phi'$, 
as shown in Figure~\ref{fig:ggfusion}. As consequence, the energetic products from $\chi_\mathrm{u}\bar\chi_\mathrm{u}$ decays can be readily detected, particularly when they occur alongside a mediator particle that carries substantial transverse momentum, providing greater sensitivity than that of searches where either $\chi_\mathrm{u}$ or $\phi'$ are considered in isolation. 

We probe the scenario where the scalar $\phi'$ has family non-universal fermion couplings, as was suggested in~\cite{Dutta2020}, and thus can address several issues with the SM. 
We focus on the $\phi^{\prime}$ decay to a pair of muons since, at experimental level, muons generally have high reconstruction and identification efficiencies, which allow for the development of relatively low $p_{\mathrm{T}}(\mu)$ triggers, and provide clean signatures to remove the copious QCD multijet SM background. A key component of this study is the development of an analysis strategy utilizing a machine learning (ML) algorithm based on Boosted Decision Trees (BDT)~\cite{friedman_greedy_2001}. The event classifier's output is employed to conduct a profile-binned likelihood test, which is used to determine the overall signal significance for each model examined in the analysis. The effectiveness of BDTs and other ML algorithms has been validated in numerous experimental and phenomenological studies \cite{,Ai:2022qvs, ATLAS:2017fak, Biswas:2018snp,  Chung:2020ysf, Feng:2021eke, ttZprime, Chigusa:2022svv,  Florez2023, Arganda2024, Ajmal_2024, Dutta_2015}. Our findings indicate that the BDT algorithm significantly enhances signal significance.

The rest of this paper is structured as follows. Section~\ref{sec:model} discuss details of the minimal  $U(1)_{T_R^3}$ model. Section~\ref{sec:exp} provides an overview of current relevant results at the LHC. Section~\ref{sec:sims} explains how the Monte
Carlo simulation samples are produced for this study. In Section~\ref{sec:ML} we discuss the motivation and details of our machine learning workflow, and in Section~\ref{sec:results}, the main results are presented. We conclude with a short discussion in Section~\ref{sec:discussion}.

\section{Experimental Considerations}\label{sec:exp}

\begin{figure}
\centering
\includegraphics[width=0.75\linewidth]{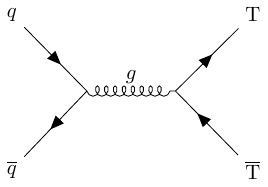}
\caption{Representative Feynman diagram for T pair production via gluon-mediated QCD processes.\label{fig:qcd_T_prod}}
\end{figure}

\begin{figure}
\centering
\includegraphics[width=0.75\linewidth]{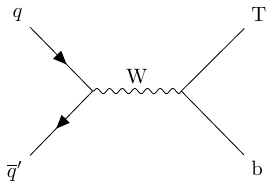}
\caption{Representative Feynman diagram for single T production via electroweak processes.\label{fig:qed_T_prod}}
\end{figure}

The ATLAS and CMS collaborations at CERN have conducted various searches for heavy vector-like quarks (T). These searches utilized $\mathrm{pp}$ collisions at center-of-mass energies of $\sqrt{s} = 8$ and $13$ \textrm{TeV}. The studies primarily focused on T production through gluon-mediated QCD processes, either in pair production from quark-antiquark annihilation (Figure~\ref{fig:qcd_T_prod}) or in single-T production from electroweak processes involving associated quarks (Figure~\ref{fig:qed_T_prod}). 

In those studies, \textrm{T} decays into $\mathrm{bW}$, $\mathrm{tZ}$, or $\mathrm{tH}$ have been considered. In the context of \textrm{T} pair production, $\mathrm{T}\bar{\mathrm{T}}$, via QCD processes, the cross sections are well-known and solely depend on the mass of the vector-like quark.  Assuming a narrow $\mathrm{T}$ decay width ($\Gamma / m(\mathrm{T}) < 0.05$ or 0.1) and a 100\% branching fraction to $\textrm{bW}$, $\textrm{tZ}$, or $\textrm{tH}$, these searches have set stringent bounds on $m(\mathrm{T})$, excluding masses below almost 1.5 \textrm{TeV} at 95\% confidence level~\cite{CMS:2024bni,CMS:2024qdd,ATLAS:2022ozf,ATLAS:2023bfh,ATLAS:2022hnn,ATLAS:2022tla,ATLAS:2023pja,ATLAS:2024fdw}. The most recent analysis from the CMS collaboration probes T-quark production via $\mathrm{pp} \to \mathrm{T}\textrm{qb}$, in final states with $\mathrm{T} \to \textrm{tZ}$ or $\mathrm{T} \to \textrm{tH}$, considering scenarios with preferential couplings to third-generation fermions. The analysis sets 95\% confidence level upper limits of 68-1260 \textrm{fb} on the production cross section, for T masses ranging from 600-1200 \textrm{GeV}~\cite{CMS:2024qdd}. The latest studies from ATLAS probe vector-like quarks using the single-T production mode with the $\mathrm{T} \to \textrm{tH}$ decay channel leading to a fully hadronic final state~\cite{ATLAS:2022ozf}, the single-T production mode with the $\mathrm{T} \to \textrm{tZ}$ decay channel leading to a multileptonic final state~\cite{ATLAS:2023bfh}, the TT pair production mode with various T decay channels leading to multileptonic final states~\cite{ATLAS:2022hnn}, and the TT pair production mode with various T decay channels leading to a single lepton plus missing momentum final state~\cite{ATLAS:2022tla,ATLAS:2023pja}. 
The multilepton search offers the greatest sensitivity in most of the phase space, but the missing transverse energy based search has better sensitivity for low branching fraction $\mathfrak{B}(\mathrm{T}\to \textrm{Wb})$ and high $\mathfrak{B}(\mathrm{T}\to \textrm{Ht})$. These searches have similar sensitivities for the singlet and doublet models, resulting in exclusion bounds for masses below about 1.25 \textrm{TeV} and 1.41 \textrm{TeV}, respectively. 

A key consideration in the model interpretations summarized above is that the $\mathrm{T}$ branching fractions depend on the chosen model. 
The excluded mass range is less restrictive for specific branching fraction scenarios, such as $\{\mathfrak{B}(\mathrm{T} \to \textrm{tZ})$, $\mathfrak{B}(\mathrm{T} \to bW)$, $\mathfrak{B}(\mathrm{T} \to \textrm{tH})\}= \{0.2, 0.6, 0.2\}$, excluding masses below about 0.95 \textrm{TeV}. Moreover, if the $\mathrm{T} \to \phi't $ decay is allowed, or if the branching fractions $\mathfrak{B}(\mathrm{T} \to \textrm{tH/bW})$ are lower, the limits previously quoted must be re-evaluated. The authors in~\cite{Cacciapaglia:2019zmj} emphasize that bounds on $m(\mathrm{T})$ can be around 500 \textrm{GeV} when $\mathrm{T} \to \mathrm{t}\phi'$ decays are permitted. Therefore, to facilitate a comprehensive study, benchmark scenarios in this paper are considered down to $m(\chi_\mathrm{u}) = 500$ \textrm{GeV}.

\section{\boldmath The Minimal $U(1)_{T_R^3}$ Model}\label{sec:model}
\subsection{Scalar Potential}
In this model, the SM group is extended by the Abelian gauge symmetry $U(1)_{T^3_R}$, where only right-handed fermions are charged. We assume two independent Higgs mechanisms, one with a Higgs doublet $\mathrm{H}$ for electroweak symmetry breaking and the other with a Higgs singlet $\phi$ for the $U(1)_{T^3_R}$ symmetry breaking. Both scalars have independent vacuum expectation values (VEVs), $\expval{H}=v_h/\sqrt2$ and $\expval\phi =v_\phi/\sqrt2$, allowing to express the doublet and singlet Higgs fields, following a Kibble parametrization, as 
\begin{align}
    H & = \begin{pmatrix}
        G_{+} \\
        \frac{1}{\sqrt{2}}\left(v_h+\rho_0+i G_{0}\right)
    \end{pmatrix}\label{eq:higgskibblepara1}
    \\
    \phi & =\frac{1}{\sqrt{2}}\left(v_\phi + \rho_\phi+i G_{\phi}\right). \label{eq:higgskibblepara2}
\end{align}
In Eqs.\ref{eq:higgskibblepara1} and Eq.\ref{eq:higgskibblepara2}, $G_\pm$, $G_0$, and $G_\phi$ are the Goldstone bosons that allow the SM $\textrm{W}^\pm$ and $\textrm{Z}$ bosons and the dark photon $A'$, associated with the $U(1)_{T^3_R}$ symmetry, to acquire mass. The $\rho_h$ and $\rho_\phi$ are an orthogonal mixture of the SM Higgs boson and the dark Higgs
\begin{equation}
    \begin{pmatrix}
        h
        \\
        \phi'
    \end{pmatrix}
    =
    \begin{pmatrix}
        \cos\alpha & -\sin\alpha
        \\
        \sin\alpha & \cos\alpha
    \end{pmatrix}
    \begin{pmatrix}
        \rho_0
        \\
        \rho_\phi
    \end{pmatrix},
\end{equation}
that result from the diagonalization of the mass matrices arising from the gauge invariant potential
\begin{equation}
    \begin{aligned}
        \mathcal V(\phi,H)
    &= \mu_H^2 H^{\dagger} H 
    +\mu_\phi^2 \phi^* \phi
    \\
    &+\lambda\left(H^{\dagger} H\right)\left(\phi^* \phi\right)
    +\lambda_H\left(H^{\dagger} H\right)^2
    +\lambda_\phi\left(\phi^* \phi\right)^2.
    \end{aligned}
\end{equation}
The tadpole equations are given from the minimization of the potential as
\begin{align}
    \pdv{\mathcal V}{H} 
     &= \frac{v_h}{\sqrt2} \left( \mu_H^2 +\lambda_Hv_h^2 + \frac{1}{2} \lambda v_\phi^2 \right) = 0,
    \\
    \pdv{\mathcal V}{\phi}
    &= \frac{v_\phi}{\sqrt2} \left( \mu_\phi^2 +\lambda_\phi v_\phi^2 + \frac{1}{2} \lambda v_h^2 \right) = 0,
\end{align}
raising the mass of the scalar bosons as
\begin{equation}
    \begin{aligned}
        m_{h,\phi'}^2 &= \frac{1}{2}\left( 
    \lambda_H v_h^2 + \lambda_\phi v_\phi^2
    \right)\\
    &\pm 
    \sqrt{
        \lambda^2 v_h^2 v_\phi^2
        +
        \left(
        \lambda_H v_h^2 - \lambda_\phi v_\phi^2
        \right)^2
    },
    \end{aligned}
\end{equation}
and the mixing angle $\alpha$ as
\begin{equation}
    \tan \alpha = \frac{-\lambda v_h v_\phi}{ \lambda_H v_h^2 - \lambda_\phi v_\phi^2 - \sqrt{\lambda^2 v_h^2 v_\phi^2 + \left(\lambda_H v_h^2 - \lambda_\phi v_\phi^2\right)^2}}.
\end{equation}
\subsection{The Universal Seesaw Mechanism}
In the model, each electrically charged SM fermion $f$ has a mass protected by both VEVs. In turn, they  acquire mass from the mixture with a vector-like fermion $\chi_f$, which is charged as the right-handed component of the respective SM fermion, in a UV complete theory. The terms in the Lagrangian density that contribute to the mass of physical fermions are,
\begin{equation}
    \begin{aligned}
        -\mathcal{L}&\supset 
    Y_{f_L} \bar{f}_L' \chi_{fR}' H 
    +Y_{f_R} \bar\chi_{fL}' f'_R  \phi^* 
    + m_{\chi_f'} \bar{\chi}_{f L}' \chi_{f R}'\\
&+\text { h.c.}
    \end{aligned}
\end{equation}
Therefore, in the vacuum, the mass matrix is
\begin{equation}
    M_f=
    \begin{pmatrix}
    0 & Y_{f_L} v_h /\sqrt2\\
    Y_{f_R} v_\phi /\sqrt2 & m_{\chi_f'}    
    \end{pmatrix}.
\end{equation}
The left- and right-handed components of the physical fermions $(f,\,\chi_f)$ are given by two rotations $\mathcal R(\theta_{f_{L,R}})$ as, 
\begin{equation}
    \begin{pmatrix}
        f_{L,R}
        \\
        \chi_{f_{L,R}}
    \end{pmatrix}
    =
    \begin{pmatrix}
        \pm\cos\theta_{f_{L,R}} & \mp \sin \theta_{f_{L,R}}
        \\
        \sin \theta_{f_{L,R}} & \cos\theta_{f_{L,R}}
    \end{pmatrix}
    \begin{pmatrix}
        f_{L,R}'
        \\
        \chi_{f_{L,R}}'
    \end{pmatrix},
\end{equation}
in a way that $\mathcal{R}(\theta_{f_L})M_f\mathcal{R}^{-1}(\theta_{f_R})=\text{diag}(m_f,m_{\chi_f})$ up to a phase. Assuming real parameters, the physical masses and the mixing angles are given by
\begin{gather}
    m_f m_{\chi_f}=\frac{ \left(Y_{f_{L}} v_h\right) \left(Y_{f_R} v_\phi\right)}{2}, \label{eq:prodmass}
     \\ 
    m_f^2 + m_{\chi_f}^2 = m_{\chi_f'}^2 + \frac{1}{2}\left(Y_{f_L}^2v_h^2+Y_{f_R}^2v_\phi^2\right),\label{eq:summass}
    \\
    \tan \theta_{f_{L,R}} =  \frac{\sqrt 2}{m_{\chi_f'}}\left(\frac{Y_{f_{L,R}}v_{h,\phi}}{2} - 
    \frac{m_f^2}{Y_{f_{L,R}}v_{h,\phi}} \right).
\end{gather}

\noindent The Yukawa interactions of the physical fermions with the scalar bosons have the form
\begin{equation}
    -\mathcal{L}_{\text{yuk}} 
    = h \bar\psi_{f_L} \mathcal{Y}_{f_L}\psi_{f_R} + \phi' \bar\psi_{f_L} \mathcal{Y}_{f_R}\psi_{f_R},
\end{equation} 
with $\psi_{f} = (f,\chi_{f})^T$, and the matrices $\mathcal{Y}_{f_{L,R}}$ given by
\begin{align}
    \mathcal{Y}_{f_L} &= \frac{1}{\sqrt{2}}
    \mathcal{R}(\theta_{f_L})
    \left(
        Y_{f_L}\sigma_+ \cos\alpha 
    - 
    Y_{f_R}\sigma_-\sin\alpha
    \right)
    \mathcal{R}^{-1}(\theta_{f_R})\label{eq:YukawaL}
    \\
    \mathcal{Y}_{f_R} &= \frac{1}{\sqrt{2}}
    \mathcal{R}(\theta_{f_L})
    \left(
    Y_{f_L}\sigma_+ \sin\alpha
    +
    Y_{f_R}\sigma_-\cos\alpha
    \right)
    \mathcal{R}^{-1}(\theta_{f_R}),\label{eq:YukawaR}
\end{align}
where $\sigma_{\pm}=(\sigma_1\pm i\sigma_2)/2$ are the ladder Pauli matrices.

\subsection{Minimal UV-complete theory}

The model must provide non-zero masses for all the SM fermions and be free of gauge anomalies. So, we must have at least one full generation of vector-like fermions $\{\chi_\mathrm{u}$, $\chi_\mathrm{d}$, $\chi_\mathrm{\ell}$, $\chi_\mathrm{\nu}\}$ and the right-handed component of the SM neutrinos, $\nu_R$, charged as shown in Table~\ref{tab:QMnumbers}. Therefore, the Yukawa interactions in the UV-complete theory must be of the form
\begin{equation}
    \begin{aligned}
        -\mathcal{L}
        \supset&\quad
        Y_{L u}^i \bar{q}_L^{\prime i} \chi_{u R}' \widetilde{H}
        + Y_{R u}^i \bar{\chi}_{u L}' u_R^{\prime i} \phi^*  
        + m_{\chi_\mathrm{u}} \bar{\chi}_{u L}' \chi_{u R}'
        \\&
        +Y_{L d}^i \bar{q}_L^{\prime i} \chi_{d R}' H 
        +Y_{R d}^i \bar{\chi}_{d L}' d_R^{\prime i} \phi
        +m_{\chi_d} \bar{\chi}_{d L}' \chi_{d R}'
        \\&
        +Y_{L \ell}^{i} \bar{\ell}_L^{\prime i} \chi_{\ell R}' H
        +Y_{R \ell}^{i} \bar{\chi}_{\ell L}' \ell_R^{\prime i} \phi
        +m_{\chi_\ell} \bar{\chi}_{\ell L}' \chi_{\ell R}'
        \\
        &
        +Y_{L \nu}^{i} \bar{\ell}_L^{\prime i} \chi_{\nu R}' \widetilde{H}
        +Y_{R \nu}^{i} \bar{\chi}_{\nu L}' \nu_R^{\prime i} \phi^*
        +m_{\chi_\nu} \bar{\chi}_{\nu L}' \chi_{\nu R}' \\
        &+\text { h.c., }
    \end{aligned}
\end{equation}
where the $i$ index runs over the three generations of fermions. The simultaneous diagonalization of the mass matrices of each fermion sector will have a similar structure to the one presented in Eqs.~\ref{eq:prodmass} and~\ref{eq:summass} and the yukawa matrices will have a similar structure of Eqs.~\ref{eq:YukawaL} and~\ref{eq:YukawaR} but codifying the $CKM$ matrix. For the neutrinos sector, the structure of the mass matrix will be more complex due to the presence of the additional Majorana mass term for the vector-like neutrino $\chi_\nu'$.

\begin{table}[h]
    \centering
    \begin{tabular}{ccccc}
    \hline
    \hline
        Field & $SU(3)_C$  & $SU(2)_L$ & $U(1)_Y$ & $U(1)_{T^3_R}$ \\
    \hline\hline
        $q_L'$                    & \bf{3} & \bf{2} & 1/6 & 0\\
        $\ell_L'$                 & \bf{1} & \bf{2} & -1/2 & 0\\
        $H$                         & \bf{1} & \bf{2} & 1/2 & 0\\
        \hline
        $u_R^{\prime c}$          & \bf{3} & \bf{1} & -2/3 & -2\\
        $d_R^{\prime c}$          & \bf{3} & \bf{1} & 1/3 & 2\\
        $\ell_R^{\prime c}$       & \bf{1} & \bf{1} & 1 & 2\\
        $\nu_R^{\prime c}$        & \bf{1} & \bf{1} & 0 & -2\\
        $\phi$                      & \bf{1} & \bf{1} & 0 & 2\\
        \hline
        $\chi_{u_L}'$               & \bf{3} & \bf{1} & 2/3 & 0\\
        $\chi_{u_R}^{\prime c}$     & \bf{3} & \bf{1} & -2/3 & 0\\
        $\chi_{d_L}'$               & \bf{3} & \bf{1} & -1/3 & 0\\
        $\chi_{d_R}^{\prime c}$     & \bf{3} & \bf{1} & 1/3 & 0\\
        $\chi_{\ell_L}'$            & \bf{1} & \bf{1} & -1 & 0\\
        $\chi_{\ell_R}^{\prime c}$  & \bf{1} & \bf{1} & 1 & 0\\
        $\chi_{\nu_L}'$             & \bf{1} & \bf{1} & 0 & 0\\
        $\chi_{\nu_R}^{\prime c}$   & \bf{1} & \bf{1} & 0 & 0\\
    \hline
    \hline
    \end{tabular}
    \caption{Minimal field content of the model and their representations under the SM and $U(1)_{T^3_R}$ gauge groups.}
    \label{tab:QMnumbers}
\end{table}

\section{Samples and Simulation}\label{sec:sims}

Samples of simulated signal and background events are generated with \texttt{MadGraph5\_aMC} (v3.2.0)~\cite{Alwall:2014hca,Alwall:2014bza} considering \textrm{pp} beams colliding with a center-of-mass energy of $\sqrt{s} = 13.6$ \textrm{TeV}. Moreover, we use the \texttt{NNPDF3.0~NLO}~\cite{NNPDF:2014otw} set for parton distribution functions (PDFs) for all event generation. Parton level events are then interfaced with \texttt{PYTHIA} (v8.2.44)~\cite{Sjostrand:2014zea} to account for parton showering and hadronization processes. Finally, we use  \texttt{DELPHES} (v3.4.2)~\cite{deFavereau:2013fsa} to simulate smearing and other detector effects using the CMS detector geometric configurations and parameters for particle identification and reconstruction, using the CMS input card with 140 average pileup interactions. All cross sections used in this analysis are obtained requiring the following kinematic criteria on leptons $\ell$, \textrm{b} quarks, and light-quark/gluon jets ($j$) at parton level in \texttt{MadGraph}: $p_{\mathrm{T}}(\ell) > 35$~\textrm{GeV}, $\abs{\eta (\mathrm{b})} < 2.5$, $\abs{\eta (\ell)} < 2.3$, $p_{\mathrm{T}}(j) > 20$~\textrm{GeV}, and $\abs{\eta (\mathrm{j})} < 5$. Furthermore, we use the MLM algorithm for jet matching and jet merging. The parameters \texttt{xqcut} and \texttt{qcut} of the MLM algorithm are set to 30 and 45 respectively to ensure continuity of the differential jet rate as a function of jet multiplicity. Each simulated signal and background sample is produced with one million events at generation level

\begin{figure}
    \centering
    \includegraphics[width=\linewidth]{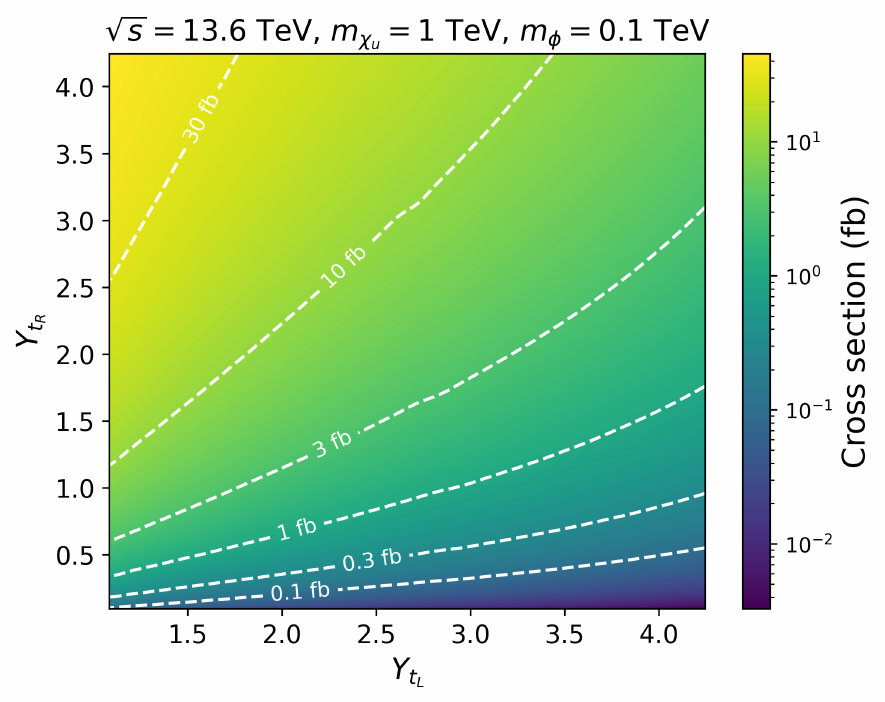}
    \caption{Signal production cross section, $ \mathrm{pp}\to \chi_\mathrm{u} \mathrm{t} \phi'$,  in the $Y_{\mathrm{t}_R}$ versus $Y_{\mathrm{t}_L}$ plane, for a benchmark point with $m(\phi')=100$~\textrm{GeV} and $m(\chi_\mathrm{u})=1.00$~\textrm{TeV}. The white-dashed contours show specific cross section values in the two dimensional plane.}
    \label{fig:cross_section_by_lambdas}
\end{figure}

Signal samples are generated considering the production of a $\phi'$ boson, an associated $\chi_\mathrm{u}$ vector-like quark, and a top quark $(\mathrm{pp}\to \chi_\mathrm{u} \mathrm{t} \phi')$, inclusive in both $\alpha$ and $\alpha_s$ (see Figures~\ref{fig:qqfusion}-\ref{fig:ggfusion}). We have used the implementation of the $U(1)_{T^3_R}$ model in Ref.~\cite{Dutta2023}. Signal samples were created considering coupling values of $Y_{\mathrm{t}_R}=Y_{\mathrm{t}_L}=2\sqrt{2}$ in the range of masses $m(\phi')\in\{5,10,50,100,325\}$~\textrm{GeV} for the dark higgs and $m(\chi_\mathrm{u})\in\{0.50, 0.75, 1.0, 1.5, 2.0, $ $ 2.5\}$~\textrm{TeV} for the vector like quark $\chi_u$~\cite{PhysRevD.108.095006}. The production cross section for $\mathrm{pp}\to \chi_\mathrm{u} \mathrm{t} \phi'$ is highly dependent on the choice of the Yukawa couplings in the Lagrangian. The ${\chi_\mathrm{u}}{- \mathrm{t}}$ fusion process shown in Figure~\ref{fig:qqfusion} is dominated by the $Y_{\mathrm{t}_R}$ coupling. However, the decay ${\chi_\mathrm{u}} \to \mathrm{t} \phi'$ shown in Figure~\ref{fig:ggfusion} is inversely proportional to the $Y_{\mathrm{t}_L}$ coupling. This effect is shown in Figure~\ref{fig:cross_section_by_lambdas}, which displays the total signal cross section, as a function of $Y_{\mathrm{t}_R}$ and $Y_{\mathrm{t}_L}$, for a benchmark point with $m(\phi')=100$~\textrm{GeV} and $m(\chi_\mathrm{u})=1.0$~\textrm{TeV}. 

\begin{figure}
    \centering
    \includegraphics[width=\linewidth]{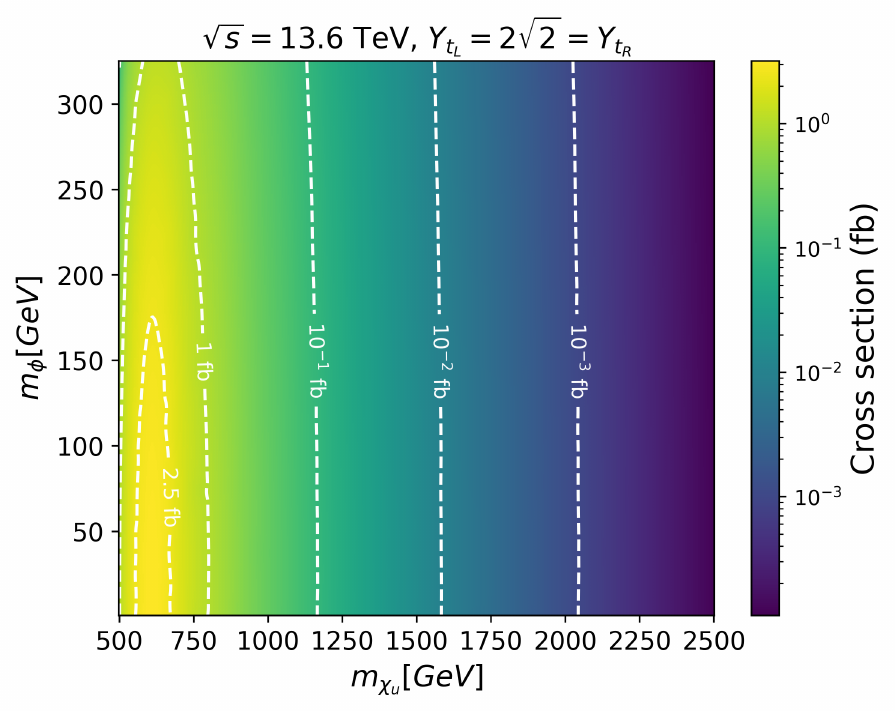}
    \caption{Projected cross section (fb) plot for $pp\to t \chi_\mathrm{u} \phi'$ and subsequent decay as a function of $m(\chi_\mathrm{u})$ and $m(\phi')$.}
    \label{fig:xs-plot}
\end{figure}

We target signal events where the top quark decays hadronically into a bottom quark and two jets ($\mathrm{t} \to \mathrm{bW} \to \mathrm{b} q \bar{q}'$), the $\chi_\mathrm{u}$ decays semileptonically into a $b$ quark, lepton, and neutrino (via $\chi_\mathrm{u} \to \mathrm{bW}$ and $\mathrm{W}\to\mu\nu_{\mu}$), and the $\phi'$ produces two muons. We note that the scalar $\phi'$ particle could result from the mixture of the SM Higgs boson and additional scalar fields, and the Yukawas of the fermions could additionally arise from the mixing of the SM fermions with additional copies of the associated vector-like fermions. Therefore, the $\phi'$ branching ratios are dependent on the chosen mechanism and model by which this mixture occurs, see for example~\cite{Cacciapaglia_2023,Blankenburg:2012nx,Jones-Perez:2013oia,Calibbi:2009pv}. For the purpose of this work, and 
similar to Refs.~\cite{Dutta2020,Dutta2023}, the considered benchmark signal scenarios have $\mathfrak{B}(\chi_\mathrm{u} \rightarrow \textrm{b W})$ of about 0.5 and $\mathfrak{B}(\phi' \rightarrow \mu^+\mu^-)=0.98$. Figure~\ref{fig:xs-plot} shows the production cross section in \textrm{fb}, as a function of $m(\phi')$ and $m(\chi_\mathrm{u})$ masses, assuming the aforementioned decays, branching ratios and couplings.

We note that for the parameter space of focus in this paper, the total mass of the $t$-$\chi_\mathrm{u}$ system is larger than $m(\phi')$, thus the large rest energy of the $t$-$\chi_\mathrm{u}$ system is converted into potentially large momentum values for the $\phi'$. Similarly, the $t$-quark produced through the $\chi_\mathrm{u}$-$t$ fusion interaction can also have large momentum values, and thus in some cases the hadronic $t$ decay products cannot be fully reconstructed independently of each other. This results in three possible $t$ reconstruction scenarios: a fully merged scenario where the $\mathrm{W}\to jj$ system and the $\mathrm{b}$ quarks are very collimated and reconstructed as a single ``fat jet’’ (henceforth referred to as a FatJet, FJ); a partially merged scenario, where the decay products of the $\mathrm{W}$ boson form a single FatJet but the $\mathrm{b}$ quark can still be separately identified; and an un-merged scenario where all decay products can be independently identified. Each scenario has an associated identification efficiency and misidentification rate, which depends on the choice of the boosted $t$/$W$ algorithm (our choice of efficiency and misidentification rates are described later). 

Based on the above details, the final state of interest in this paper consists of three muons (two from the $\phi'$ decay and one from the $\chi_\mathrm{u}$ decay), a (possibly boosted) top-tagged system, at least one $b$-tagged jet, and large missing transverse momentum ($\vec{p}_{T}^{\textrm{~miss}}$). For the partially merged and un-merged scenarios, there will be two $b$ quarks present in the final state (one of which is part of the top tagged system). 

We consider background sources from SM processes which can give similar objects in the final state as those expected for signal. Several background sources were considered and studied, such as QCD multijet events, production of vector boson pairs ($\mathrm{VV: WW, ZZ, WZ}$), vector boson triplets ($\mathrm{VVV: WWZ, WZZ, ZZZ, WWW}$), top-quark pairs in association with weak bosons ($\mathrm{t}\overline{\mathrm{t}}X$), and $\mathrm{t}\overline{\mathrm{t}}\mathrm{t}\overline{\mathrm{t}}$ processes. The  dominant sources of SM backgrounds events are from the $\mathrm{t}\overline{\mathrm{t}}X$, $\mathrm{ZZW}$, and $\mathrm{t}\overline{\mathrm{t}}\mathrm{t}\overline{\mathrm{t}}$ processes. The $\mathrm{t}\overline{\mathrm{t}}X$ background is primarily associated production of a $\mathrm{Z}/\gamma^{*}$ from $\mathrm{t}\bar{\mathrm{t}}$ fusion processes. The $\mathrm{ZZW}$ process becomes a background when one $\mathrm{Z}$ decays $\mathrm{b}\bar{\mathrm{b}}$, another $\mathrm{Z}$ decays to a pair of muons, and the W decays to a muon and a neutrino. 
Events from $\mathrm{ZZW}$ and $\mathrm{t}\overline{\mathrm{t}}\mathrm{t}\overline{\mathrm{t}}$ have been combined, after being weighted by their corresponding production cross section. The combination is presented as the ``$\mathrm{b} \overline{\mathrm{b}}\mu\mu\mu\nu$'' background in the remainder of this paper. The $\mathrm{t}\overline{\mathrm{t}}X$ process is presented as part of the ``$\mathrm{t}\overline{\mathrm{t}}\mu^{+}\mu^{-}$'' background. Table~\ref{tab:dominantbkgs} shows the production cross sections for the dominant background sources. The rest of the aforementioned background processes do not contribute meaningfully in our context, accounting for $\ll 1\%$ of the total expected background yield.

The identification of leptons, boosted top quarks, and bottom quarks plays an important role the ability to identify signal events, the ability to minimize the rate of SM backgrounds, and thus also the discovery reach in the high-luminosity environment of the LHC. It is worth noting that the reconstruction and identification of leptons and the decay products of the top/bottom quarks may be non-trivial at the High-Luminosity LHC (HL-LHC) due to the presence of a potentially large number of secondary pp interactions (pileup). The impact of pileup on the new physics discovery reach, and the importance of pileup mitigation at CMS and ATLAS has been outlined in many papers, for example in Ref.~\cite{CMS-PAS-FTR-13-014}. We note the expected performance of the upgraded ATLAS and CMS detectors for the HL-LHC is beyond the scope of this work, however, the studies presented here do attempt to provide reasonable expectations by conservatively assuming some degradation in lepton and hadron identification efficiencies, using Ref.~\cite{CMS-PAS-FTR-13-014} as a benchmark, and considering the case of 140 average pileup interactions. 

For muons with $|\eta|< 1.5$, the assumed identification efficiency is 95\% with a 0.3\% misidentification rate~\cite{CMS-PAS-FTR-13-014,CMS_MUON_17001}. The performance degrades linearly with $\eta$ for $1.5 < |\eta| < 2.5$, and we assume an identification efficiency of 65\% with a 0.5\% misidentification rate at $|\eta| = 2.5$. Similarly, the charged hadron tracking efficiency, which contributes to the jet clustering algorithm and missing transverse momentum ($\vec{p}_{T}^{\textrm{~miss}}$) calculation, is 97\% for $1.5 < |\eta| < 2.5$, and degrades to about 85\% at $|\eta| = 2.5$. These potential inefficiencies due to the presence of secondary pp interactions contribute to how well the lepton and top kinematics can be reconstructed. Following Refs.~\cite{CMS:2020poo,ATLAS:2018wis}, we consider the ``Loose'' working point for the identification of the fully merged (partially merged) $\mathrm{t}$ decays, which results in 80-85\% top (W) identification efficiency and 11-25\% misidentification rate, depending on the FatJet transverse momentum ($p_{T}^{FJ}$). Following Ref.~\cite{CMSbtag}, we consider the ``Loose'' working point of the DeepCSV algorithm~\cite{Bols_2020}, which gives a 70-80\% b-tagging efficiency and 10\% light quark mis-identification rate. The choice of boosted $t$/$W$ and b-tagging working points is determined through an optimization process which maximizes discovery reach. It is noted the contribution from SM backgrounds with a misidentified boosted $t$/$W$ is negligible, and thus our discovery projections are not sensitive to uncertainties related to the boosted $t$/$W$ misidentification rates. 

\begin{table}[]
  \begin{tabular}{l r}
    \hline
    {Background Process} & {Cross-Section $\sigma$ [\textrm{pb}]} \\
    \hline
   $\mathrm{pp} \to \mathrm{t} \overline{\mathrm{t}} \, \mu^+ \mu^-$ & $2.574\times 10^{-3}$  \\
    $\mathrm{pp} \to \mathrm{b}\overline{\mathrm{b}}\, \mu\mu\mu\nu $ & $4.692 \times 10^{-4}$ \\
    \hline
  \end{tabular}
  \centering
  \caption{A summary of dominant SM backgrounds produced by $\mathrm{pp}$ collisions and their cross sections in pb, as computed by \texttt{MadGraph} with $n = 10^6$ events.}
  \label{tab:dominantbkgs}
\end{table}

\begin{figure}[]
\centering
\includegraphics[width=\linewidth]{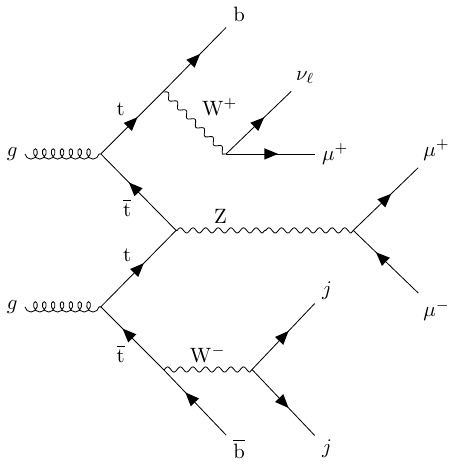}
\caption{Representative Feynman diagram for a background event. A $Z$ boson is produced in association with a top quark through the fusion of a top, anti top pair from incoming protons. The $Z$ boson subsequently decays to a pair of muons and the two spectator top quarks decay semi-leptonically and purely hadronically to muons, neutrinos and jets, resulting in the same final states as the signal event.\label{fig:v}}
\end{figure}

\section{Data Analysis Using Machine Learning}\label{sec:ML}
The analysis of signal and background events is performed utilizing machine learning techniques. A machine learning-based approach offers sizeable advantages when compared to traditional event classification techniques. Unlike conventional methods, machine learning models have the capability to simultaneously consider all kinematic variables, allowing them to efficiently navigate the complex and high-dimensional space of event kinematics. Consequently, machine learning models can effectively enact sophisticated selection criteria that take into account the entirety of this high-dimensional space. This makes them ideal for high energy physics applications.

The BDT method is a powerful machine learning technique that has proven its effectiveness in various applications, particularly in the field of collider physics. In this method, decision trees are trained greedily in a sequential manner, with each tree focusing on learning the discrepancies or residuals between its predictions and the expected values obtained from the previously trained tree. This iterative process aims to progressively minimize errors, making BDTs a particularly effective approach for enhancing model performance.

In the context of collider physics, BDTs have demonstrated their utility in addressing classification problems. In particular, BDTs can effectively discriminate between signal and background events, enabling accurate and efficient event classification. Their ability to handle subtle non-linear relationships within the data with high interpretability makes BDTs a valuable tool to handle large amounts of data with a large number of parameters for each event. 

The first step in our workflow involves the use of a specialized \texttt{MadAnalysis Expert Mode} C++ script~\cite{CONTE2013222}. This script extracts essential kinematic and topological information from the simulated samples. The script will process the aforementioned variables contained within these files and transform them into a structured and informative CSV (Comma-Separated Values) format that can be used to train our machine learning models. These kinematic variables include crucial details about the events, such as particle momenta, energies, and topologies, providing the fundamental building blocks for our machine learning analysis. Figure~\ref{fig:feature_importance} show the features that are used for training the machine learning models and their importance for a benchmark point.

To account for the differential significance of various events, we apply cross-section weighting. This ensures that the relative importance of signal and background events is appropriately balanced in the dataset. This weighting is crucial for addressing the varying likelihood of observing different types of events in high-energy physics experiments. The prepared and weighted datasets are then passed to our \texttt{MadAnalysis Expert Mode} C++ script, where the simulated signal and background events are initially filtered, before being passed to the CSV file for use by the machine learning algorithm. The filtering process requires at least one well reconstructed and identified $\mathrm{b}$-jet candidate, at least one jet (regular or FJ) not tagged as a $\mathrm{b}$ jet, and exactly three identified muons. The filtering selections are motivated by experimental constraints, such as the geometric constraints of the CMS/ATLAS detectors, the typical kinematic thresholds for reconstruction of particle objects, and the available lepton triggers which also drive the minimal kinematic thresholds. Selected jets must have $p_{\mathrm{T}} > 30$ $\textrm{GeV}$ and $|\eta(j)| < 5.0$, while $\mathrm{b}$-jet candidates with $p_{\mathrm{T}} > 20$ $\textrm{GeV}$ and $|\eta(\mathrm{b})| < 2.5$ are chosen. The $\mu$ object must pass a $p_{\mathrm{T}} > 35$ $\textrm{GeV}$ threshold and be within a $|\eta(\ell)| < 2.3$. We will refer to this filtering criteria as pre-selections. The efficiency of the pre-selections depends on $m(\phi')$ and $m(\chi_{\mathrm{u}})$, but is typically about 25-30\% for the signal samples. Events passing this pre-selection are used as input for the machine learning algorithm, which classifies them as signal or background, using a probability factor. 

\begin{figure}
\centering
\includegraphics[width=\linewidth]{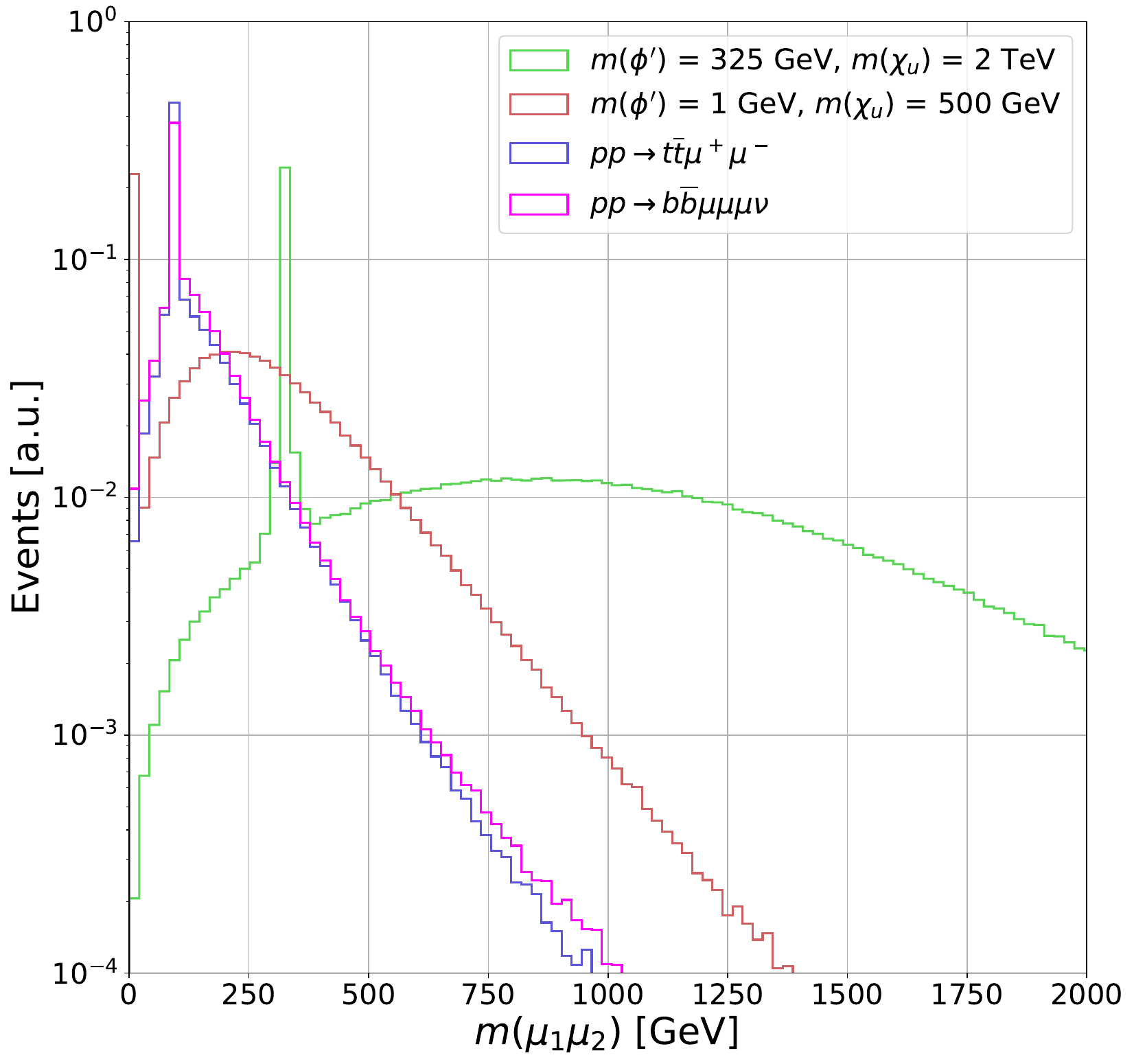}
\caption{Invariant mass distribution of the muon pair with the highest and second highest transverse momentum. The distributions are shown for the two main SM background processes and two signal benchmark points.\label{fig:_mu12}}
\end{figure}

We explore the performance of a diverse set of machine learning models, specifically three neural networks of differing architectures and a BDT algorithm. To ensure robust model assessment, we employed a standard 90-10 train-test split of the dataset, partitioning it into a 90\% portion for training and a 10\% portion for testing. This division allows us to gauge the generalization capabilities of our models on unseen data.  

The training and evaluation of the BDT was carried out in a high-performance computing environment. Specifically, an Nvidia A100 GPU was used. The canonical \texttt{PyTorch}~\cite{paszke2019} deep learning framework was employed for configuring, training, and evaluating the neural networks. PyTorch is well-regarded for its flexibility and performance in deep learning applications.

For the BDT algorithm, we used hyperparameters $\eta=0.3$, $\gamma = 0$, and $\texttt{max\_depth} = 6$. The \texttt{XGBoost}~\cite{chen_xgboost_2016} library was used for the implementation of the Boosted Decision Tree algorithm. It offers high efficiency, optimization, and interpretability, making it a suitable choice for this particular task. 

\begin{figure}
\centering
\includegraphics[width=\linewidth]{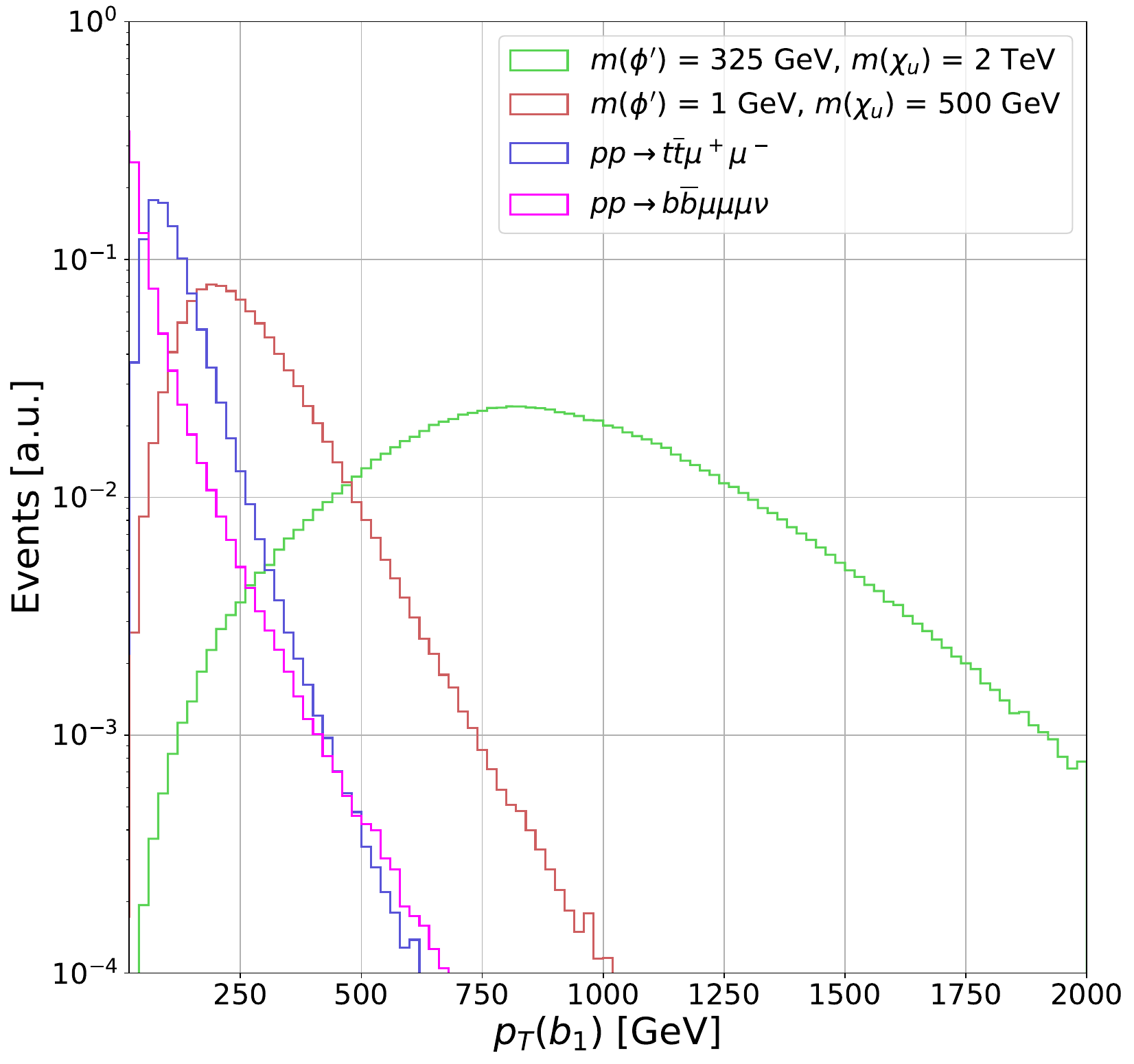}
\caption{Transverse momentum distribution of the leading \textrm{b}-quark jet candidate. The distributions are shown for the two main SM background processes and two signal benchmark points.\label{fig:pTb1}}
\end{figure}

\begin{figure}
\centering
\includegraphics[width=\linewidth]{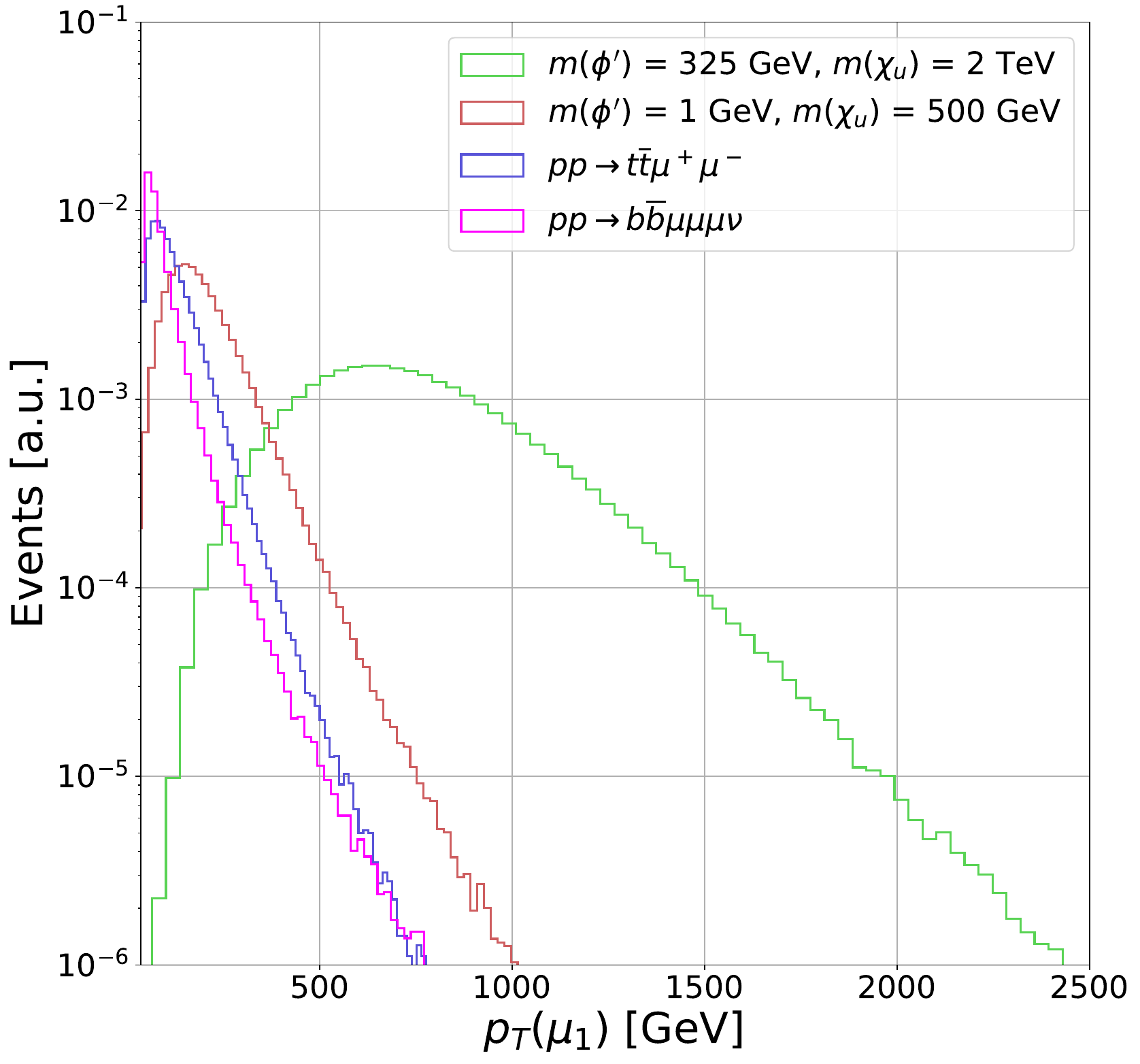}
\caption{Transverse momentum distribution of the leading muon candidate. The distributions are shown for the two main SM background processes and two signal benchmark points.\label{fig:pTmu1}}
\end{figure}

\begin{table}
    \centering
    \begin{tabular}{c  c  c}
    \hline
    { Model} & { Train/Test Acc. } & { Training Time} \\
    \hline
    \small
    BDT & N.A./0.9993  & 6s\\
    Neural Network 1 & 0.9999/0.9997 & 1h 58m \\
    Neural Network 2 & 0.9999/0.9998 & 2h 12m \\
    Neural Network 3 & 0.9999/0.9998 & 2h 32m\\
    \hline
    \end{tabular}
    \caption{Train/test results for the ML models.}
    \centering
\end{table}

It is worth mentioning that we experimented with deep neural networks of various architectures. Although we found that they yield similar signal sensitivity to the BDT, the complex nature of the studies in this
work (particle objects considered, experimental constraints in a high luminosity LHC, etc.) motivate the use of a BDT over a deep neural network because of its usefulness, efficiency, and simplicity understanding the machine learning output in addition to significantly shorter training times. Therefore, we perform our proceeding analysis using the BDT. The outcomes of our model training and evaluation are presented in Table 3. 

\section{Results}\label{sec:results}
\begin{figure}
\centering
  \centering
  \includegraphics[width=\linewidth]{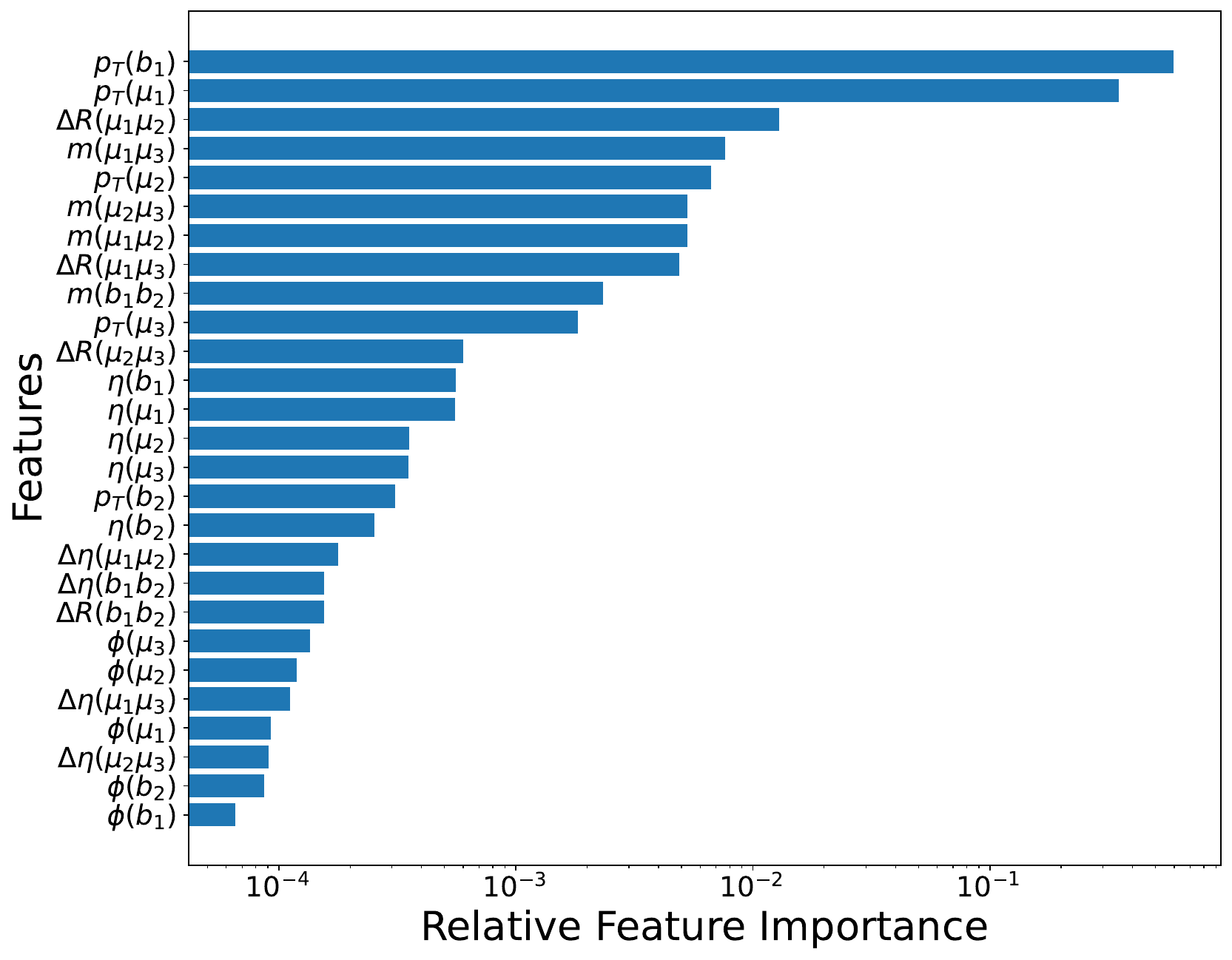}
  \caption{Relative importance of features in training for a benchmark signal scenario with $m(\phi')=325\, \mathrm{GeV}$ and $m(\chi_\mathrm{u})=2000\, \mathrm{GeV}$.}
  \label{fig:feature_importance}
\end{figure}

Figures~\ref{fig:_mu12},~\ref{fig:pTb1}, and~\ref{fig:pTmu1}, show relevant kinematic distributions for two benchmark signal points and the dominant SM backgrounds, using the subset of events passing the pre-selections defined above. The signal benchmark points in these figures are $m(\phi^{'}) = 325 \, \mathrm{GeV}$, $m(\chi_{\mathrm{u}}) = 2\, \mathrm{TeV}$, and $m(\phi^{'}) = 1 \, \mathrm{GeV}$, $m(\chi_{\mathrm{u}}) = 500\, \mathrm{GeV}$. The distributions are normalized such that the area under the curve is unity. These distributions correspond to the reconstructed mass, $m(\mu_{1}, \mu_{2})$, between the two muon candidates with the highest transverse momentum ($\mu_{1}$ and $\mu_{2}$), 
the transverse momentum of the b-jet candidate with the highest transverse momentum $p_{\mathrm{T}}$ ($\mathrm{b_{1}}$), and the muon candidate with the highest transverse momentum $p_{\mathrm{T}}$ ($\mu_{1}$), respectively. 
These distributions are among the variables identified by the BDT algorithm with the highest signal to background discrimination power (see Figure~\ref{fig:feature_importance}).

As can be seen from Figure~\ref{fig:_mu12}, the $\phi'$ mass can be reconstructed through its associated muon decay pair, which is observed as a peak in the $m(\mu_{1}, \mu_{2})$ distribution around the expected $m(\phi')$ value, and has low- and high-mass tails which are a consequence of cases where the leading and/or subleading muon is not from the $\phi'$ decay, but rather from the associated $\mathrm{W}$ boson from the $\chi_{\mathrm{u}}$ decay. For the backgrounds, muons come from \textrm{Z} (\textrm{W}) decays. Therefore, the $m(\mu_{1}, \mu_{2})$ background distributions show a peak near $m_{\mathrm{W/Z}}$, combined with a broad distribution indicative of the combination of two muon candidates from different decay vertices. We note that the $\phi'\to\mu^{+}\mu^{-}$ decay width depends on the square of the $\phi'\to\mu^{+}\mu^{-}$ coupling and  $\frac{m_{\mu}^{2}}{m(\phi')^{2}}$ and is thus suppressed by the relatively small muon mass. Therefore, the width of the $m(\mu_{1}, \mu_{2})$ signal distributions is driven by the experimental resolution in the reconstruction of the muon momenta, as well as the probability that the two leading muons are the correct pair from the $\phi'$ decay. Since the probability that the two highest-$p_{\mathrm{T}}$ muons are the correct pair from the $\phi'\to\mu^{+}\mu^{-}$ decay depends on $m(\phi')$ and $m(\chi_\mathrm{u})$, it is important to include all possible combinations of dimuon pairs (i.e., $m(\mu_{1}, \mu_{3})$ and $m(\mu_{2}, \mu_{3})$) in the training of the BDT. 

Figure~\ref{fig:pTb1} shows the  distribution for the \textrm{b}-jet candidate with the highest $p_{\mathrm{T}}$, $p_{\mathrm{T}}(\mathrm{b}_1)$, for the same simulated samples shown in Figure~\ref{fig:_mu12}. Based on the signal topology and our choise of parameter space (i.e., $m(\chi_\mathrm{u}) > m_{\mathrm{t}}$), it is expected that the leading $\mathrm{b}$-jet candidate come from the $\chi_\mathrm{u}$ decay, with an average $p_{\mathrm{T}}$ close to $\frac{m(\chi_\mathrm{u}) - m_{\mathrm{W}}}{2}$, as observed in Figure~\ref{fig:pTb1}. For the $\mathrm{t} \overline{\mathrm{t}} \mu^{+}\mu^{-}$ background, the \textrm{b}-jet candidates come from top-quark decays. Therefore, their average transverse momentum is expected to be $\frac{m_{\mathrm{t}} - m_{\mathrm{W}}}{2} \approx 45$ GeV, as observed in Figure~\ref{fig:pTb1}. On the other hand, the \textrm{b}-jet candidates for the $\mathrm{b} \overline{\mathrm{b}}\mu\mu\mu\nu$ background can come from off-mass shell $\mathrm{Z}^{*}/\gamma^{*}$, and thus typically have an even softer spectrum in comparison to the $\mathrm{t} \overline{\mathrm{t}} \mu^{+}\mu^{-}$ background.

Figure~\ref{fig:pTmu1} shows the  distribution for the muon candidate with the highest $p_{\mathrm{T}}$, $p_{\mathrm{T}}(\mu_{1})$. Similar to Figure~\ref{fig:pTb1}, when $m(\chi_\mathrm{u}) > m_{\mathrm{t}}$ it is expected that the leading muon candidate come from the $\chi_\mathrm{u}$ decay, with an average $p_{\mathrm{T}}$ of approximately $\frac{m(\chi_\mathrm{u}) - m_{\mathrm{W}}}{4}$, as observed in Figure~\ref{fig:pTmu1}. For the major SM backgrounds, the muon candidates come from Z/W/$\gamma^{*}$ decays. Therefore, their average transverse momentum is expected to be much lower, $\frac{m_{\mathrm{Z/W}}}{4} \approx 40-45$ GeV. This kinematic feature provides a nice handle to discriminate high $m(\chi_\mathrm{u})$ signal events amongst the large SM backgrounds, which have lower average $p_{\textrm{T}}(\mu)$ constrained by the SM weak boson masses.

In addition to these aforementioned variables in Figures~\ref{fig:_mu12}-\ref{fig:pTmu1}, several other kinematic variables were included as inputs to the BDT algorithm. In particular, 27 such variables were used in total, 
and these included the momenta of $\mathrm{b}$ and muon candidates; invariant masses of pairs of muons; angular differences between 
$\mathrm{b}$ jets and between the muons. 
As mentioned above, the variables $m(\mu_{i}, \mu_{j})$ for $i, j \neq 1$ provide some additional discrimination between signal and background when the leading muons are not a $\phi'$ decay candidate. The angular separation variables, such as $\Delta R(\mu_{i}, \mu_{j})$, are designed to be sensitive to lower mass $\phi'$, since the low rest mass of those particles means they acquire more boost, and thus smaller angular separation $\Delta R$ between the muon candidates. The trained BDT returns the discriminating power of each of its inputs, and the feature importance for each variable is shown in Figure~\ref{fig:feature_importance} for a signal benchmark point with $m(\phi')=325\, \mathrm{GeV}$ and $m(\chi_\mathrm{u})=2000\, \mathrm{GeV}$.

\begin{figure}
\centering
  \centering  \includegraphics[width=\linewidth]{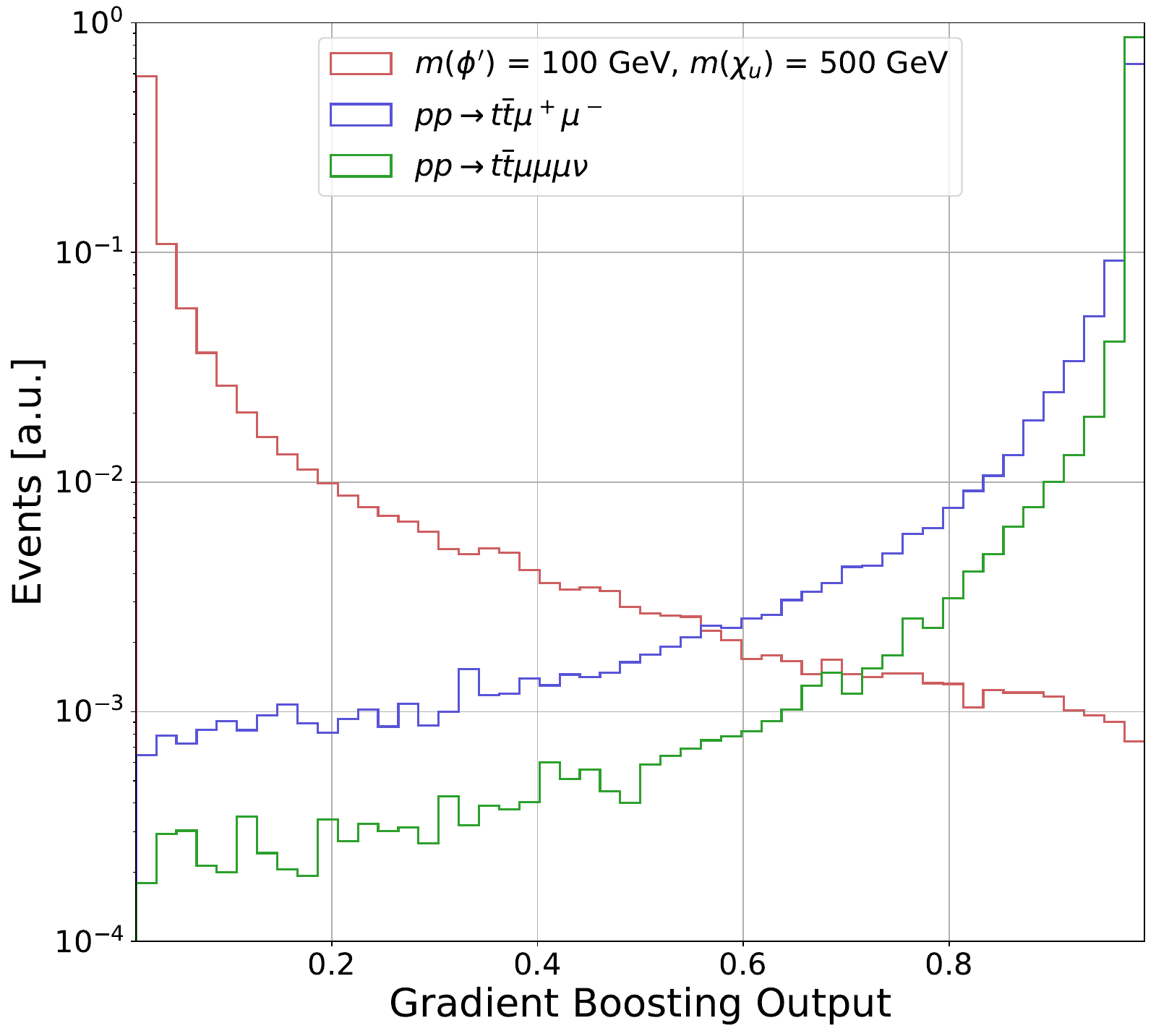}
  \caption{Output of the gradient boosting algorithm for a benchmark $m(\phi') = 100$ GeV and $m(\chi_\mathrm{u}) = 500\, \mathrm{GeV}$ signal, and dominant backgrounds. The distributions are normalized to unity.}
  \label{fig:xgboostout}
\end{figure}

\begin{figure}
\centering
  \centering
  \includegraphics[width=\linewidth]{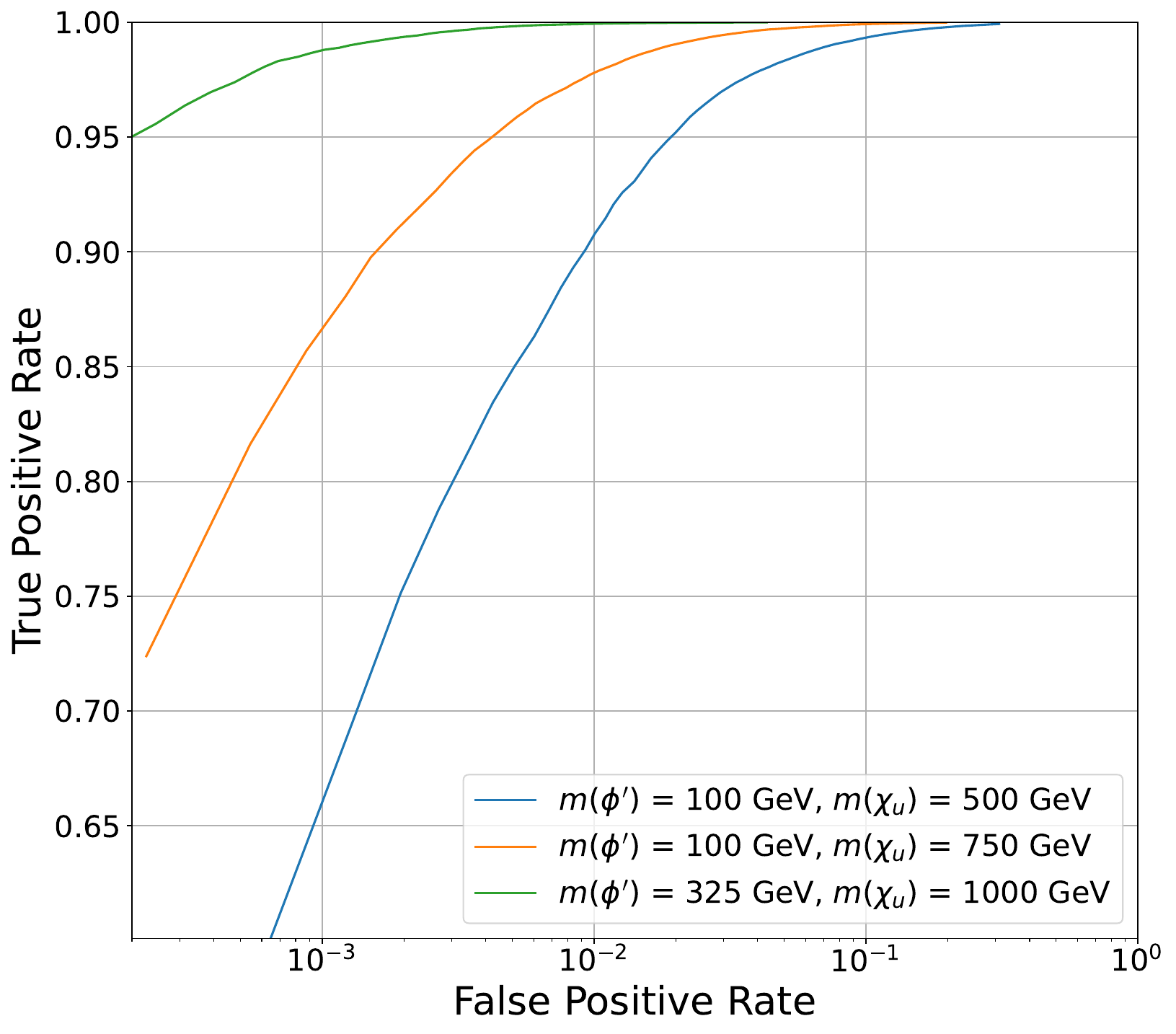}
  \caption{Receiver operating characteristic curve of the BDT algorithm for three different signal benchmark scenarios.}
  \label{fig:ROC}
\end{figure}

Figure~\ref{fig:xgboostout} shows  the distributions for the output of the BDT algorithm, normalized to unity, for the representative signal benchmark point of $m(\phi') = 1\, \mathrm{GeV}$, $m(\chi_\mathrm{u}) = 0.5\, \mathrm{TeV}$ and the two dominant backgrounds. The output of the BDT algorithm is a value between 0 and 1, which quantifies the likelihood that an event is either background-like (BDT output near 1) or signal-like (BDT output near 0). Figure~\ref{fig:ROC} illustrates the true positive rate (TPR), defined as the probability of correctly selecting signal events using the BDT output, plotted against the false positive rate (FPR), defined as the probability of incorrectly selecting background events. For example, for $m(\phi') = 100\, \mathrm{GeV}$ and $m(\chi_\mathrm{u}) = 500\, \mathrm{GeV}$, when signal events are selected at 65\% probability, the background is selected at about $10^{-3}$ probability.

The outputs from the BDT machine learning algorithm are used to perform a profile-bin likelihood analysis to estimate the signal significance for a luminosity of 3000 $\mathrm{fb^{-1}}$, corresponding to the expected amount of collected data by the end of the LHC era. For this purpose, the BDT distributions are normalized to cross section times pre-selection efficiency times luminosity for the different signal models. The significance is then calculated using the expected bin-by-bin yields of the BDT output distribution in a profile likelihood fit, using the ROOTFit \cite{Butterworth:2015oua} package developed by CERN. The expected signal significance $Z_\text{sig}$ is calculated using the probability of obtaining the same test statistic for the  signal plus background and the signal-null hypotheses, defined as the local $p$-value. Similar to Refs. \cite{Florez:2021zoo, Florez:2019tqr, Florez:2018ojp, Florez:2017xhf, VBFZprimePaper, Florez:2016lwi, Leonardi_2020}, the significance  corresponds to the point where the integral of a Gaussian distribution between $Z_\text{sig}$ and $\infty$ results in a value equal to the local $p$-value. The estimation of $Z_\text{sig}$ incorporates  systematic uncertainties. The uncertainty values have been included as as nuisance parameters, considering lognormal priors for normalization and Gaussian priors for uncertainties associated with the modeling of the shapes similar to Refs. \cite{natalia2021longtermlhcdiscoveryreach, PhysRevD.103.095001}. 

The systematic uncertainties that have been included result from experimental and theoretical constraints.   A 1-5\% systematic uncertainty, depending on the simulated MC sample, has been included to account for the choice of Parton Distribution Function (PDF) set. The systematic uncertainty effect was incorporated following the PDF4LHC~\cite{Butterworth:2015oua} recommendations. This systematic uncertainty has a small impact on the expected event yields for signal and background, but it does not affect the shape of the BDT output distribution. We additionally considered theoretical uncertainties related to the absence of higher-order contributions to the signal cross sections, which can change the pre-selection efficienciencies and the shapes of kinematic variables used as inputs to the BDT algorithm. This uncertainty was calculated by varying the renormalization and factorization scales by $\times 2$, and studying the resulting change in the bin-by-bin yields of the BDT distributions. They are found to be at most 2\% in a given bin. 

Regarding experimental uncertainties, following experimental measurements from CMS on the estimation of the integrated luminosity, a conservative 3\% effect has been included~\cite{lumiRef}. A 5\% systematic uncertainty associated with the reconstruction and identification of $\mathrm{b}$-quark jets has been included, independent of $p_\mathrm{T}$ and $\eta$ of the $\mathrm{b}$-jet candidates. According to Ref.~\cite{CMSbtag}, this uncertainty is correlated between signal and background processes with genuine  \textrm{b}-jets and is also correlated across BDT bins for each process. For muons, we include a 2\% uncertainty associated with the reconstruction, identification, and isolation requirements, and a 3\% systematic uncertainty to account for scale and resolution effects on the momentum and energy measurement. 
We consider jet energy scale uncertainties ranging from 2-5\%, contingent on $\eta$ and $p_\mathrm{T}$, resulting in shape-based uncertainties on the BDT output distribution. Jet energy scale uncertainties were assumed to range from 1-5\%, contingent on $\eta$ and $p_\mathrm{T}$. These assumptions lead to shape-based uncertainties on the BDT output distribution, varying from 1-2\%. Additionally, we include  a 10\% systematic uncertainty to account for errors in the signal and background predictions. Considering all the various sources of systematic uncertainties our conservative  estimate yields a total effect of about 20\%.

\begin{figure}[]
\centering
  \centering
  \includegraphics[width=\linewidth]{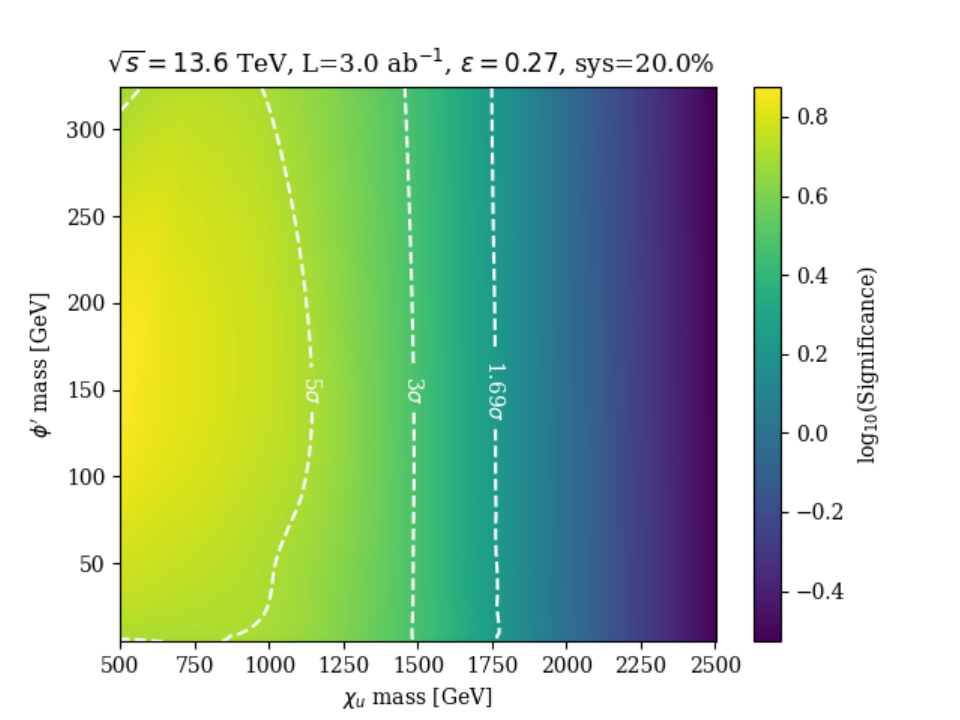}
  \caption{Signal significance for the high luminosity LHC era, considering with 3000  $\mathrm{fb}^{-1}$ of collected data.}
  \label{fig:/significance_3000}
\end{figure}

Figure~\ref{fig:/significance_3000} shows the expected signal significance considering an integrated luminosity of 3000 $\mathrm{fb^{-1}}$. The significance is shown as a heat map in a two dimensional plane for different $\phi'$ and $\chi_{\mathrm{u}}$ masses. The x-axis corresponds to $m(\chi_\mathrm{u})$, the y-axis to $m(\phi')$, and the heat map to log$_{10}(\mathrm{Z}_{sig})$. The white dashed lines are contours of constant signal signifances of $1.69 \sigma$,  $3\sigma$ and  $5\sigma$ to represent regions of possible exclusion, evidence of new physics, and discovery, respectively. Under these conditions, $\phi'$ ($\chi_{\mathrm{u}}$) masses ranging from 1 to 325 \textrm{GeV} (500 to 1800 \textrm{GeV}) can be probed. The range for a discovery with $5\sigma$ signal significance varies from $\chi_{\mathrm{u}}$ masses from $m(\chi_{\mathrm{u}}) = 770$-1100 \textrm{GeV}, depending  $m(\phi^{'})$. 


\section{Discussion}\label{sec:discussion}

The LHC will continue to run with pp collisions at $\sqrt{s} = 13.6$ TeV for the next decade. Given the increase in the integrated luminosity expected from the high-luminosity program, it is important to consider unexplored new physics phase space that diverges from the conventional assumptions made in many BSM theories, and which could have remained hidden in processes that have not yet been thoroughly examined. It is additionally crucial to explore advanced analysis techniques, in particular the use of artifical intelligence algorithms, to enhance the probability of detecting these rare corners where production cross sections are lower and discrimination from SM backgrounds is difficult. 

In this work, we examine a model based on a $U(1)_{T^3_R}$ extension of the SM, which can address various conceptual and experimental issues with the SM, including the mass hierarchy between generations of fermions, the thermal dark matter abundance, and the muon $g - 2$, $R_{(D)}$, and $R_{(D^*)}$ anomalies. This model contains a light scalar boson $\phi'$, with potential masses below the electroweak scale, and TeV-scale vector-like quarks $\chi_\mathrm{u}$. We consider the scenario where the scalar $\phi'$ has family non-universal fermion couplings and $m(\phi') \ge 1$ GeV, as was suggested in~\cite{Dutta2020}, and thus the $\phi^{\prime}$ can primarily decay to a pair of muons. Previous works in Refs.~\cite{Dutta2023, Banerjee_2016} considered scenarios motivating a search methodology with a merged diphoton system from $\phi' \to \gamma\gamma$ decays. The authors of Ref~\cite{Dutta2023}, in which $m(\phi') < 1$ GeV,  indeed pointed out that if the $\phi'$ is heavier than about 1 GeV, then decays to $\mu^+ \mu^-$ can become the preferable mode for discovery, which is the basis for the work presented in this paper. We further note that the final state topology studied in this paper would represent the most important mode for discovery at $m(\phi') < 2 m_{\mathrm{t}}$ where the $\phi' \to \mathrm{t\bar{t}}$ decay is kinematically forbidden. 

The main result of this paper is that we have shown that the LHC can probe the visible decays of new bosons with masses below the electroweak scale, down to the GeV-scale, by considering the simultaneous production of heavy QCD-coupled particles, which then decay to the SM particles that contain large momentum values and can be observed in the central regions of the CMS and ATLAS detectors. The boosted system combined with innovative machine learning algorithms allows for the signal extraction above the lower-energy SM background. The LHC search strategy described here can be used to discover the prompt decay of new light particles.  An important conclusion from this paper is that the detection prospects for low-mass particles are enhanced when it is kinematically possible to simultaneously access the heavy degrees of freedom which arise in the UV completion of the low-energy model.  This specific scenario in which the couplings of the light scalars are generationally dependent, with important coupling values to the top quark, is an ideal example which would be difficult to directly probe at low energy beam experiments.

The proposed data analysis represents a competitive alternative 
to complement searches already being conducted at the LHC, allowing us to probe $\phi'$ masses from 1 to 325 \textrm{GeV}, for $m(\chi_{\mathrm{u}})$ values up to almost 2 TeV, at the HL-LHC. Therefore, we strongly encourage the ATLAS and CMS Collaborations to consider the proposed analysis strategy in future new physics searches. 

\begin{acknowledgements}
The authors would like to thank Joel Jones-Pérez for fruitful discussions.
A.F and C.R thank the constant and enduring financial support received for this project from the faculty of science at Universidad de Los Andes (Bogot\'a, Colombia) through the projects INV-2023-178-2999 and INV-2023-175-2957. 
A.G and U.Q acknowledge the funding received from the Physics \& Astronomy department at Vanderbilt University and the US National Science Foundation. 
This work is supported in part by NSF Award PHY-1945366 and a Vanderbilt Seeding Success Grant.

\end{acknowledgements}

\bibliographystyle{spphys}       
\bibliography{refs}   

\begin{thebibliography}{100}
\providecommand{\url}[1]{{#1}}
\providecommand{\urlprefix}{URL }
\expandafter\ifx\csname urlstyle\endcsname\relax
  \providecommand{\doi}[1]{DOI \discretionary{}{}{}#1}\else
  \providecommand{\doi}{DOI \discretionary{}{}{}\begingroup \urlstyle{rm}\Url}\fi

\bibitem{PhysRevD.73.072003}
G.W. Bennett, et~al., Phys. Rev. D \textbf{73}, 072003 (2006).
\newblock \doi{10.1103/PhysRevD.73.072003}

\bibitem{g2cit}
M.~Tanabashi, et~al., Phys. Rev. D \textbf{98}, 030001 (2018).
\newblock \doi{10.1103/PhysRevD.98.030001}

\bibitem{Davier2017}
M.~Davier, A.~Hoecker, B.~Malaescu, Z.~Zhang, The European Physical Journal C \textbf{77}(12), 827 (2017).
\newblock \doi{10.1140/epjc/s10052-017-5161-6}

\bibitem{Davier2020}
M.~Davier, A.~Hoecker, B.~Malaescu, Z.~Zhang, The European Physical Journal C \textbf{80}(3), 241 (2020).
\newblock \doi{10.1140/epjc/s10052-020-7792-2}

\bibitem{PhysRevLett.121.022003}
T.~Blum, P.A. Boyle, V.~G\"ulpers, T.~Izubuchi, L.~Jin, C.~Jung, A.~J\"uttner, C.~Lehner, A.~Portelli, J.T. Tsang, Phys. Rev. Lett. \textbf{121}, 022003 (2018).
\newblock \doi{10.1103/PhysRevLett.121.022003}

\bibitem{PhysRevD.97.114025}
A.~Keshavarzi, D.~Nomura, T.~Teubner, Phys. Rev. D \textbf{97}, 114025 (2018).
\newblock \doi{10.1103/PhysRevD.97.114025}

\bibitem{PhysRevLett.124.132002}
T.~Blum, N.~Christ, M.~Hayakawa, T.~Izubuchi, L.~Jin, C.~Jung, C.~Lehner, Phys. Rev. Lett. \textbf{124}, 132002 (2020).
\newblock \doi{10.1103/PhysRevLett.124.132002}

\bibitem{PhysRevD.100.076004}
F.~Campanario, H.~Czy\ifmmode~\dot{z}\else \.{z}\fi{}, J.~Gluza, T.~Jeli\ifmmode~\acute{n}\else \'{n}\fi{}ski, G.~Rodrigo, S.~Tracz, D.~Zhuridov, Phys. Rev. D \textbf{100}, 076004 (2019).
\newblock \doi{10.1103/PhysRevD.100.076004}

\bibitem{BaBar:2012obs}
B.~Collaboration, Phys. Rev. Lett. \textbf{109}, 101802 (2012).
\newblock \doi{10.1103/PhysRevLett.109.101802}

\bibitem{BaBar:2013mob}
B.~Collaboration, Phys. Rev. D \textbf{88}(7), 072012 (2013).
\newblock \doi{10.1103/PhysRevD.88.072012}

\bibitem{Huschle:2015rga}
{Belle Collaboration}, Phys. Rev. D \textbf{92}, 072014 (2015).
\newblock \doi{10.1103/PhysRevD.92.072014}

\bibitem{LHCb:2015gmp}
L.~Collaboration, Phys. Rev. Lett. \textbf{115}(11), 111803 (2015).
\newblock \doi{10.1103/PhysRevLett.115.111803}.
\newblock [Erratum: Phys.Rev.Lett. 115, 159901 (2015)]

\bibitem{Aaij:2015yra}
{LHCb Collaboration}, Phys. Rev. Lett. \textbf{115}, 111803 (2015).
\newblock \doi{10.1103/PhysRevLett.115.111803}.
\newblock [Erratum: Phys.Rev.Lett. 115, 159901 (2015)]

\bibitem{Sato:2016svk}
{Belle Collaboration}, Phys. Rev. D \textbf{94}, 072007 (2016).
\newblock \doi{10.1103/PhysRevD.94.072007}

\bibitem{Hirose:2016wfn}
{Belle Collaboration}, Phys. Rev. Lett. \textbf{118}, 211801 (2017).
\newblock \doi{10.1103/PhysRevLett.118.211801}

\bibitem{Aaij:2017uff}
{LHCb Collaboration}, Phys. Rev. Lett. \textbf{120}, 171802 (2018).
\newblock \doi{10.1103/PhysRevLett.120.171802}

\bibitem{Hirose:2017dxl}
{Belle Collaboration}, Phys. Rev. D \textbf{97}, 012004 (2018).
\newblock \doi{10.1103/PhysRevD.97.012004}

\bibitem{LHCb:2017rln}
L.~Collaboration, Phys. Rev. D \textbf{97}(7), 072013 (2018).
\newblock \doi{10.1103/PhysRevD.97.072013}

\bibitem{Abdesselam:2019dgh}
{Belle Collaboration},   (2019).
\newblock \doi{10.48550/arXiv.1904.08794}

\bibitem{Belle:2019rba}
{Belle Collaboration}, Phys. Rev. Lett. \textbf{124}, 161803 (2020).
\newblock \doi{10.1103/PhysRevLett.124.161803}

\bibitem{LHCb:2023zxo}
L.~Collaboration,   (2023).
\newblock \doi{10.48550/arXiv.2302.02886}

\bibitem{ParticleDataGroup:2024cfk}
S.~Navas, et~al., Phys. Rev. D \textbf{110}(3), 030001 (2024).
\newblock \doi{10.1103/PhysRevD.110.030001}

\bibitem{CMS:2018iye}
A.M. Sirunyan, et~al., JHEP \textbf{03}, 170 (2019).
\newblock \doi{10.1007/JHEP03(2019)170}

\bibitem{CMS:2016ucr}
V.~Khachatryan, et~al., Phys. Rev. Lett. \textbf{118}(2), 021802 (2017).
\newblock \doi{10.1103/PhysRevLett.118.021802}

\bibitem{CMS:2016xbv}
V.~Khachatryan, et~al., JHEP \textbf{02}, 048 (2017).
\newblock \doi{10.1007/JHEP02(2017)048}

\bibitem{CMS:2016fxb}
V.~Khachatryan, et~al., JHEP \textbf{03}, 077 (2017).
\newblock \doi{10.1007/JHEP03(2017)077}

\bibitem{CMS:2017xcw}
A.M. Sirunyan, et~al., JHEP \textbf{07}, 121 (2017).
\newblock \doi{10.1007/JHEP07(2017)121}

\bibitem{CMS:2015jsu}
V.~Khachatryan, et~al., JHEP \textbf{11}, 189 (2015).
\newblock \doi{10.1007/JHEP11(2015)189}

\bibitem{PatiSalam1974}
J.C. Pati, A.~Salam, Physical Review D \textbf{10}(1), 275 (1974).
\newblock \doi{10.1103/PhysRevD.10.275}

\bibitem{MohapatraPati1975}
R.N. Mohapatra, J.C. Pati, Physical Review D \textbf{11}(9), 2558 (1975).
\newblock \doi{10.1103/PhysRevD.11.2558}

\bibitem{SenjanovicMohapatra1975}
G.~Senjanovic, R.N. Mohapatra, Physical Review D \textbf{12}(5), 1502 (1975).
\newblock \doi{10.1103/PhysRevD.12.1502}

\bibitem{DiLuzio2018}
L.~Di~Luzio, J.~Fuentes-Martin, A.~Greljo, M.~Nardecchia, S.~Renner, Journal of High Energy Physics \textbf{2018}(11), 81 (2018).
\newblock \doi{10.1007/JHEP11(2018)081}

\bibitem{Baker2019}
M.J. Baker, J.~Fuentes-Martín, G.~Isidori, M.~König, The European Physical Journal C \textbf{79}(4), 334 (2019).
\newblock \doi{10.1140/epjc/s10052-019-6853-x}

\bibitem{Michaels:2020fzj}
L.~Michaels, F.~Yu, JHEP \textbf{03}, 120 (2021).
\newblock \doi{10.1007/JHEP03(2021)120}

\bibitem{Dev:2021otb}
P.S.B. Dev, W.~Rodejohann, X.J. Xu, Y.~Zhang, JHEP \textbf{06}, 039 (2021).
\newblock \doi{10.1007/JHEP06(2021)039}

\bibitem{Florez2023}
A.~Flórez, J.~Jones-Pérez, A.~Gurrola, C.~Rodriguez, J.~Peñuela-Parra, The European Physical Journal C \textbf{83}(11), 1023 (2023).
\newblock \doi{10.1140/epjc/s10052-023-12177-4}

\bibitem{Dutta:2022qvn}
B.~Dutta, S.~Ghosh, J.~Kumar, in \emph{{Snowmass 2021}} (2022)

\bibitem{Dutta2019}
B.~Dutta, S.~Ghosh, J.~Kumar, Physical Review D \textbf{100}(7), 075028 (2019).
\newblock \doi{10.1103/PhysRevD.100.075028}

\bibitem{Dutta2020}
B.~Dutta, S.~Ghosh, J.~Kumar, Physical Review D \textbf{102}(7), 075041 (2020).
\newblock \doi{10.1103/PhysRevD.102.075041}

\bibitem{Dutta2020b}
B.~Dutta, S.~Ghosh, J.~Kumar, Physical Review D \textbf{102}(1), 015013 (2020).
\newblock \doi{10.1103/PhysRevD.102.015013}

\bibitem{Dutta2022}
B.~Dutta, S.~Ghosh, P.~Huang, J.~Kumar, Physical Review D \textbf{105}(1), 015011 (2022).
\newblock \doi{10.1103/PhysRevD.105.015011}

\bibitem{PhysRevD.107.095019}
V.~De~Romeri, J.~Nava, M.~Puerta, A.~Vicente, Phys. Rev. D \textbf{107}, 095019 (2023).
\newblock \doi{10.1103/PhysRevD.107.095019}

\bibitem{Dutta2023}
B.~Dutta, S.~Ghosh, A.~Gurrola, D.~Julson, T.~Kamon, J.~Kumar, Journal of High Energy Physics \textbf{2023}(3), 164 (2023).
\newblock \doi{10.1007/JHEP03(2023)164}

\bibitem{Berezhiani}
Z.~Berezhiani, Physics Letters B \textbf{129}(1), 99 (1983).
\newblock \doi{10.1016/0370-2693(83)90737-2}

\bibitem{Chang1987}
D.~Chang, R.N. Mohapatra, Phys. Rev. Lett. \textbf{58}, 1600 (1987).
\newblock \doi{10.1103/PhysRevLett.58.1600}

\bibitem{Davidson1987}
A.~Davidson, K.C. Wali, Phys. Rev. Lett. \textbf{59}, 393 (1987).
\newblock \doi{10.1103/PhysRevLett.59.393}

\bibitem{Rajpoot1987}
S.~Rajpoot, Modern Physics Letters A \textbf{02}(05), 307 (1987).
\newblock \doi{10.1142/S0217732387000422}

\bibitem{Babu1989}
K.S. Babu, R.N. Mohapatra, Phys. Rev. Lett. \textbf{62}, 1079 (1989).
\newblock \doi{10.1103/PhysRevLett.62.1079}

\bibitem{Babu1990}
K.S. Babu, R.N. Mohapatra, Phys. Rev. D \textbf{41}, 1286 (1990).
\newblock \doi{10.1103/PhysRevD.41.1286}

\bibitem{Bhardwaj_2022}
A.~Bhardwaj, T.~Mandal, S.~Mitra, C.~Neeraj, Physical Review D \textbf{106}(9) (2022).
\newblock \doi{10.1103/physrevd.106.095014}.
\newblock \urlprefix\url{http://dx.doi.org/10.1103/PhysRevD.106.095014}

\bibitem{Bhardwaj_2022_2}
A.~Bhardwaj, K.~Bhide, T.~Mandal, S.~Mitra, C.~Neeraj, Physical Review D \textbf{106}(7) (2022).
\newblock \doi{10.1103/physrevd.106.075024}.
\newblock \urlprefix\url{http://dx.doi.org/10.1103/PhysRevD.106.075024}

\bibitem{Bardhan_2023}
J.~Bardhan, T.~Mandal, S.~Mitra, C.~Neeraj, Physical Review D \textbf{107}(11) (2023).
\newblock \doi{10.1103/physrevd.107.115001}.
\newblock \urlprefix\url{http://dx.doi.org/10.1103/PhysRevD.107.115001}

\bibitem{Banerjee_2016}
S.~Banerjee, D.~Barducci, G.~Bélanger, C.~Delaunay, Journal of High Energy Physics \textbf{2016}(11) (2016).
\newblock \doi{10.1007/jhep11(2016)154}.
\newblock \urlprefix\url{http://dx.doi.org/10.1007/JHEP11(2016)154}

\bibitem{Alves_2024}
J.M. Alves, G.~Branco, A.~Cherchiglia, C.~Nishi, J.~Penedo, P.M. Pereira, M.~Rebelo, J.~Silva-Marcos, Physics Reports \textbf{1057}, 1–69 (2024).
\newblock \doi{10.1016/j.physrep.2023.12.004}.
\newblock \urlprefix\url{http://dx.doi.org/10.1016/j.physrep.2023.12.004}

\bibitem{friedman_greedy_2001}
J.H. Friedman, The Annals of Statistics \textbf{29}(5), 1189 (2001).
\newblock \doi{10.1214/aos/1013203451}.
\newblock Publisher: Institute of Mathematical Statistics

\bibitem{Ai:2022qvs}
X.~Ai, S.C. Hsu, K.~Li, C.T. Lu, JPCS \textbf{2438}(1), 012114 (2023).
\newblock \doi{10.1088/1742-6596/2438/1/012114}

\bibitem{ATLAS:2017fak}
A.~Collaboration, Phys. Rev. D \textbf{97}(7), 072016 (2018).
\newblock \doi{10.1103/PhysRevD.97.072016}

\bibitem{Biswas:2018snp}
A.~Biswas, D.~Kumar-Ghosh, N.~Ghosh, A.~Shaw, A.K. Swain, J. Phys. G \textbf{47}(4), 045005 (2020).
\newblock \doi{10.1088/1361-6471/ab6948}

\bibitem{Chung:2020ysf}
Y.L. Chung, S.C. Hsu, B.~Nachman, JINST \textbf{16}, P07002 (2021).
\newblock \doi{10.1088/1748-0221/16/07/P07002}

\bibitem{Feng:2021eke}
J.~Feng, M.~Li, Q.S. Yan, Y.P. Zeng, H.H. Zhang, Y.~Zhang, Z.~Zhao, JHEP \textbf{2022}(9) (2022).
\newblock \doi{10.1007/jhep09(2022)141}

\bibitem{ttZprime}
D.~Barbosa, F.~Díaz, L.~Quintero, A.~Flórez, M.~Sanchez, A.~Gurrola, E.~Sheridan, F.~Romeo, EPJC \textbf{83}(413) (2023).
\newblock \doi{10.1140/epjc/s10052-023-11506-x}

\bibitem{Chigusa:2022svv}
S.~Chigusa, S.~Li, Y.~Nakai, W.~Zhang, Y.~Zhang, J.~Zheng, Phys. Lett. B \textbf{833}, 137301 (2022).
\newblock \doi{10.1016/j.physletb.2022.137301}

\bibitem{Arganda2024}
E.~Arganda, D.A. D\'{\i}az, A.D. Perez, R.M. Sand\'a~Seoane, A.~Szynkman, Phys. Rev. D \textbf{109}, 055032 (2024).
\newblock \doi{10.1103/PhysRevD.109.055032}

\bibitem{Ajmal_2024}
S.~Ajmal, J.T. Gaglione, A.~Gurrola, O.~Panella, M.~Presilla, F.~Romeo, H.~Sun, S.S. Xue, Journal of High Energy Physics \textbf{2024}(8) (2024).
\newblock \doi{10.1007/jhep08(2024)176}.
\newblock \urlprefix\url{http://dx.doi.org/10.1007/JHEP08(2024)176}

\bibitem{Dutta_2015}
B.~Dutta, A.~Gurrola, K.~Hatakeyama, W.~Johns, T.~Kamon, P.~Sheldon, K.~Sinha, S.~Wu, Z.~Wu, Physical Review D \textbf{92}(9) (2015).
\newblock \doi{10.1103/physrevd.92.095009}.
\newblock \urlprefix\url{http://dx.doi.org/10.1103/PhysRevD.92.095009}

\bibitem{CMS:2024bni}
C.~Collaboration.
\newblock Review of searches for vector-like quarks, vector-like leptons, and heavy neutral leptons in proton-proton collisions at $\sqrt{s}$ = 13 tev at the cms experiment (2024).
\newblock \urlprefix\url{https://arxiv.org/abs/2405.17605}

\bibitem{CMS:2024qdd}
C.~Collaboration.
\newblock Search for production of a single vector-like quark decaying to th or tz in the all-hadronic final state in pp collisions at $\sqrt{s}$ = 13 tev (2024).
\newblock \urlprefix\url{https://arxiv.org/abs/2405.05071}

\bibitem{ATLAS:2022ozf}
G.~Aad, et~al., Phys. Rev. D \textbf{105}(9), 092012 (2022).
\newblock \doi{10.1103/PhysRevD.105.092012}

\bibitem{ATLAS:2023bfh}
G.~Aad, et~al., Phys. Rev. D \textbf{109}(11), 112012 (2024).
\newblock \doi{10.1103/PhysRevD.109.112012}

\bibitem{ATLAS:2022hnn}
G.~Aad, et~al., Phys. Lett. B \textbf{843}, 138019 (2023).
\newblock \doi{10.1016/j.physletb.2023.138019}

\bibitem{ATLAS:2022tla}
G.~Aad, et~al., Eur. Phys. J. C \textbf{83}(8), 719 (2023).
\newblock \doi{10.1140/epjc/s10052-023-11790-7}

\bibitem{ATLAS:2023pja}
G.~Aad, et~al., JHEP \textbf{08}, 153 (2023).
\newblock \doi{10.1007/JHEP08(2023)153}

\bibitem{ATLAS:2024fdw}
G.~Aad, et~al.,   (2024)

\bibitem{Cacciapaglia:2019zmj}
G.~Cacciapaglia, T.~Flacke, M.~Park, M.~Zhang, Phys. Lett. B \textbf{798}, 135015 (2019).
\newblock \doi{10.1016/j.physletb.2019.135015}

\bibitem{Alwall:2014hca}
J.~Alwall, R.~Frederix, S.~Frixione, V.~Hirschi, F.~Maltoni, O.~Mattelaer, H.S. Shao, T.~Stelzer, P.~Torrielli, M.~Zaro, JHEP \textbf{07}, 079 (2014).
\newblock \doi{10.1007/JHEP07(2014)079}

\bibitem{Alwall:2014bza}
J.~Alwall, C.~Duhr, B.~Fuks, O.~Mattelaer, D.G. \"Ozt\"urk, C.H. Shen, Comput. Phys. Commun. \textbf{197}, 312 (2015).
\newblock \doi{10.1016/j.cpc.2015.08.031}

\bibitem{NNPDF:2014otw}
N.~Collaboration, JHEP \textbf{04}, 040 (2015).
\newblock \doi{10.1007/JHEP04(2015)040}

\bibitem{Sjostrand:2014zea}
T.~Sj\"ostrand, S.~Ask, J.R. Christiansen, R.~Corke, N.~Desai, P.~Ilten, S.~Mrenna, S.~Prestel, C.O. Rasmussen, P.Z. Skands, Comput. Phys. Commun. \textbf{191}, 159 (2015).
\newblock \doi{10.1016/j.cpc.2015.01.024}

\bibitem{deFavereau:2013fsa}
D.. Collaboration, JHEP \textbf{02}, 057 (2014).
\newblock \doi{10.1007/JHEP02(2014)057}

\bibitem{PhysRevD.108.095006}
A.C. Canbay, O.~Cakir, Phys. Rev. D \textbf{108}, 095006 (2023).
\newblock \doi{10.1103/PhysRevD.108.095006}

\bibitem{Cacciapaglia_2023}
G.~Cacciapaglia, A.~Deandrea, S.~Vatani, Physical Review D \textbf{108}(1) (2023).
\newblock \doi{10.1103/physrevd.108.016010}

\bibitem{Blankenburg:2012nx}
G.~Blankenburg, G.~Isidori, J.~Jones-Perez, Eur. Phys. J. C \textbf{72}, 2126 (2012).
\newblock \doi{10.1140/epjc/s10052-012-2126-7}

\bibitem{Jones-Perez:2013oia}
J.~Jones-Perez, J. Phys. Conf. Ser. \textbf{447}, 012060 (2013).
\newblock \doi{10.1088/1742-6596/447/1/012060}

\bibitem{Calibbi:2009pv}
L.~Calibbi, J.~Jones-Perez, A.~Masiero, J.h. Park, W.~Porod, O.~Vives, PoS \textbf{EPS-HEP2009}, 167 (2009).
\newblock \doi{10.22323/1.084.0167}

\bibitem{CMS-PAS-FTR-13-014}
\emph{{Study of the Discovery Reach in Searches for Supersymmetry at CMS with 3000/fb}}.
\newblock Geneva (2013).
\newblock \urlprefix\url{https://cds.cern.ch/record/1607141}.
\newblock CMS-PAS-FTR-13-014

\bibitem{CMS_MUON_17001}
A.~Sirunyan, et~al., JINST \textbf{15}(02), P02027 (2020).
\newblock \doi{10.1088/1748-0221/15/02/P02027}

\bibitem{CMS:2020poo}
A.M. Sirunyan, et~al., JINST \textbf{15}(06), P06005 (2020).
\newblock \doi{10.1088/1748-0221/15/06/P06005}

\bibitem{ATLAS:2018wis}
M.~Aaboud, et~al., Eur. Phys. J. C \textbf{79}(5), 375 (2019).
\newblock \doi{10.1140/epjc/s10052-019-6847-8}

\bibitem{CMSbtag}
A.~Sirunyan, et~al., JINST \textbf{13}, P05011 (2018).
\newblock \doi{10.1088/1748-0221/13/05/P05011}

\bibitem{Bols_2020}
E.~Bols, J.~Kieseler, M.~Verzetti, M.~Stoye, A.~Stakia, Journal of Instrumentation \textbf{15}(12), P12012 (2020).
\newblock \doi{10.1088/1748-0221/15/12/P12012}

\bibitem{CONTE2013222}
E.~Conte, B.~Fuks, G.~Serret, Computer Physics Communications \textbf{184}(1), 222 (2013).
\newblock \doi{10.1016/j.cpc.2012.09.009}

\bibitem{paszke2019}
A.~Paszke, et~al.
\newblock Pytorch: An imperative style, high-performance deep learning library (2019).
\newblock \urlprefix\url{https://arxiv.org/abs/1912.01703}

\bibitem{chen_xgboost_2016}
T.~Chen, C.~Guestrin, in \emph{Proceedings of the 22nd {ACM} {SIGKDD} {International} {Conference} on {Knowledge} {Discovery} and {Data} {Mining}} (Association for Computing Machinery, New York, NY, USA, 2016), {KDD} '16, pp. 785--794.
\newblock \doi{10.1145/2939672.2939785}

\bibitem{Butterworth:2015oua}
{J. Butterworth, S. Carrazza, A. Cooper-Sarkar, A. De Roeck, J. Feltesse, S. Forte, J. Gao, S. Glazov, J. Huston, Z. Kassabov, R. McNulty, A. Morsch, P. Nadolsky, V. Radescu, J. Rojo and R. Thorne}, J. Phys. G \textbf{43}, 023001 (2016).
\newblock \doi{10.1088/0954-3899/43/2/023001}

\bibitem{Florez:2021zoo}
A.~Fl\'orez, A.~Gurrola, W.~Johns, P.~Sheldon, E.~Sheridan, K.~Sinha, B.~Soubasis, Phys. Rev. D \textbf{103}(9), 095001 (2021).
\newblock \doi{10.1103/PhysRevD.103.095001}

\bibitem{Florez:2019tqr}
A.~Fl\'orez, A.~Gurrola, W.~Johns, J.~Maruri, P.~Sheldon, K.~Sinha, S.R. Starko, Phys. Rev. D \textbf{100}(1), 016017 (2019).
\newblock \doi{10.1103/PhysRevD.100.016017}

\bibitem{Florez:2018ojp}
A.~Fl\'orez, Y.~Guo, A.~Gurrola, W.~Johns, O.~Ray, P.~Sheldon, S.~Starko, Phys. Rev. D \textbf{99}(3), 035034 (2019).
\newblock \doi{10.1103/PhysRevD.99.035034}

\bibitem{Florez:2017xhf}
A.~Fl\'orez, K.~Gui, A.~Gurrola, C.~Pati\~no, D.~Restrepo, Phys. Lett. B \textbf{778}, 94 (2018).
\newblock \doi{10.1016/j.physletb.2018.01.009}

\bibitem{VBFZprimePaper}
A.~Florez, A.~Gurrola, W.~Johns, Y.~Do~Oh, P.~Shendon, D.~Teague, T.~Weiler, Phys. Lett. B \textbf{767}, 126 (2017).
\newblock \doi{10.1016/j.physletb.2017.01.062}

\bibitem{Florez:2016lwi}
A.~Fl\'orez, L.~Bravo, A.~Gurrola, C.~\'Avila, M.~Segura, P.~Sheldon, W.~Johns, Phys. Rev. D \textbf{94}(7), 073007 (2016).
\newblock \doi{10.1103/PhysRevD.94.073007}

\bibitem{Leonardi_2020}
R.~Leonardi, O.~Panella, F.~Romeo, A.~Gurrola, H.~Sun, S.S. Xue, The European Physical Journal C \textbf{80}(4) (2020).
\newblock \doi{10.1140/epjc/s10052-020-7822-0}.
\newblock \urlprefix\url{http://dx.doi.org/10.1140/epjc/s10052-020-7822-0}

\bibitem{natalia2021longtermlhcdiscoveryreach}
C.~Natalia, F.~Andrés, G.~Alfredo, J.~Will, S.~Paul, T.~cheng.
\newblock Long-term lhc discovery reach for compressed higgsino-like models using vbf processes (2021).
\newblock \urlprefix\url{https://arxiv.org/abs/2102.10194}

\bibitem{PhysRevD.103.095001}
A.~Fl\'orez, A.~Gurrola, W.~Johns, P.~Sheldon, E.~Sheridan, K.~Sinha, B.~Soubasis, Phys. Rev. D \textbf{103}, 095001 (2021).
\newblock \doi{10.1103/PhysRevD.103.095001}.
\newblock \urlprefix\url{https://link.aps.org/doi/10.1103/PhysRevD.103.095001}

\bibitem{lumiRef}
P.~Lujan, et~al., {The Pixel Luminosity Telescope: A detector for luminosity measurement at CMS using silicon pixel sensors}.
\newblock Tech. Rep. CMS-DN-21-008 (2022).
\newblock Accepted by Eur. Phys. J. C.

\end{thebibliography}



%

\end{document}